\documentclass[a4paper,11pt]{article}
\usepackage{jheppub}
\usepackage{amsmath, amssymb}
\usepackage{physics}
\usepackage{tikz}
\usetikzlibrary{arrows.meta, calc, patterns, decorations.pathmorphing}
\usepackage{xcolor}
\usepackage{bm}
\usepackage{tcolorbox}
\usepackage[subrefformat=parens]{subcaption}
\tcbset{colframe=black, colback=white, boxrule=0.8pt, arc=0pt}
\title{\boldmath 
New analytic solutions of non-SUSY black branes in $10$-D heterotic supergravity and the phase transition
}

\author{Seiken Chikazawa and Koji Hashimoto}
\affiliation{Department of Physics, Kyoto University,\\
Kyoto 606-8502, Japan}

\emailAdd{chikazawa@gauge.scphys.kyoto-u.ac.jp}
\emailAdd{koji@scphys.kyoto-u.ac.jp}

\abstract{
In the exploration of the vast string landscape, we provide new analytic solutions of non-SUSY black branes in $10$-D heterotic supergravity which exhibit a phase transition. 
Our black brane solutions carry two kinds of gauge charges but still are exact, which extends the recent developments in classification of heterotic branes.
The solutions offer a novel method to embed previously known noncritical 2-D or 3-D solutions to the critical dimensions.
Furthermore, by performing a perturbative analysis of the non-Abelian gauge field on this exact background, we analytically find that as the temperature decreases, a spontaneous symmetry breaking occurs, leading to a dynamical instability accompanied by gauge field condensation.
This instability embodies a phase transition of black branes in heterotic supergravity, and analytically predicts the existence of unknown black hole solutions to be unveiled.
Our results suggest that phase transition phenomena can be used to explore and analyze the phase structure of black hole solutions, for systematically discovering and classifying stringy black holes.}

\begin{document}
\begin{flushright}
KUNS-3085
\end{flushright}
\maketitle
\flushbottom

\section{Introduction}
Non-supersymmetric black holes often exhibit dynamical instabilities, potentially describing transitions toward more stable phases (\cite{Gregory:1993vy, Gregory:1994bj, Kol:2004ww, Harmark:2007md, Gubser:2000ec, Gubser:2000mm, Gubser:2004dr, Reall:2001ag, Straumann:1990as, Brodbeck:1994vu, Zhou:1992sb, Straumann:1989tf}).
Therefore, elucidating their properties is of great importance for understanding the stability and phase structure of spacetime in string theory.
In particular, in heterotic string theory with large rank gauge groups such as $E_8 \times E_8$ or $SO(32)$, there is room to introduce multiple types of gauge fields, leading to a vast variety of possible black hole solutions.
Understanding how these solutions transition between different phases from the viewpoint of stability is a highly intriguing problem.
However, examples of exact black hole solutions with multiple gauge fields are still limited, and it has generally been considered difficult to analytically evaluate their instabilities and trace the associated dynamics.
The reason is straightforward: it is typically impossible to analytically solve the nonlinear coupled equations involving the metric, gauge fields, and the dilaton.
When one includes several distinct gauge fields, we usually have no other effective method than numerical computation.
Even in that case, for black hole solutions, technical difficulties arise when performing numerical analysis\footnote{This is because one must solve the dynamical equations while simultaneously satisfying the constraint equations inherent in gravitational systems. Typically, one searches for black hole solutions by imposing the condition that the $(t,t)$ component of the metric $G_{tt}$ vanishes on the horizon, but then $G^{tt}$ diverges there. The validity of numerical solutions is checked by verifying whether the Hamiltonian constraint $H = 0$ holds numerically; however, when quantities like $G^{tt}$ diverge near the horizon, numerical errors cause $H$ to significantly deviate from zero, making it difficult to determine whether the constraint is truly satisfied and whether the numerical solution is genuine.}.
As a result, discussions of instabilities and phase transitions based on numerical solutions inevitably involve uncertainties. Therefore, constructing exact solutions is indispensable for rigorously understanding the spacetime dynamics.
The goal of this study is therefore to construct new exact black hole (or black brane) solutions with multiple gauge fields in heterotic supergravity and to analyze their instabilities, thereby clarifying part of the dynamics and phase structure of black holes in heterotic string theory.

Recent developments in the search for new black hole solutions in heterotic string theory have been motivated by the completeness hypothesis \cite{Polchinski:2003bq, Banks:2010zn} and the cobordism conjecture \cite{McNamara:2019rup}, which have led to the construction of new types of branes in heterotic string theory.
The completeness hypothesis, which asserts that every possible charge in quantum gravity must be carried by some physical object, allows one to predict the existence of various charged branes within string theory.
Furthermore, this hypothesis can be refined to include more subtle types of charges, leading to the cobordism conjecture, which predicts the existence of physical objects realizing any topologically nontrivial configuration at spatial infinity.
In fact, in heterotic string theory, Kaidi, Ohmori, Tachikawa, and Yonekura proposed branes with nontrivial gauge configurations \cite{1, Kaidi:2023tqo}.
These branes are constructed by taking $n$ of the ten dimensions as an $S^n$ and placing a  nontrivial gauge field on $S^n$.
This construction generalizes the heterotic black 6-brane solution originally presented by Garfinkle, Horowitz, and Strominger \cite{Garfinkle:1990qj} for $n=2$, in which the angular part of a $U(1)$ Dirac monopole is placed on $S^2$ to generate charge.
In the work \cite{1} by Kaidi, Tachikawa, and Yonekura, new cases with $n=1, 4, 8$ were considered, and the associated gauge configurations and exact CFT descriptions near the horizon throat region were discussed.
Subsequently, Fukuda, Kobayashi, Watanabe, and Yonekura (FKWY) \cite{3} realized the proposed branes as explicit solutions of heterotic supergravity, the low-energy effective theory of heterotic string theory.
There, numerical solutions for black branes with $n = 4, 8$ were constructed, and analytic solutions were derived for the compactified cases known as the extremal limits\footnote{Analytic solutions for $n=9$ were also obtained, although the physical interpretation has not been clarified.}.
See also some related work \cite{Abbott:2008cd, 2, Hellerman:2006ff, Hellerman:2007zz, Kaidi:2020jla, BoyleSmith:2023xkd, Polchinski:2005bg, Bergshoeff:2006bs, Basile:2023knk, Debray:2023rlx, Kneissl:2024zox, Montero:2020icj}.

Starting from this framework, in this paper, 
we extend the black $p$-brane FKWY solutions \cite{3} in heterotic supergravity, and construct new exact solutions by introducing temporal components of a gauge field. Previous studies related to the cobordism conjecture have focused only on configurations where gauge fields on $S^n$ generate the brane charge.
In contrast, in this work we introduce a time component $A_t(r)$ of the gauge field that depends on the radial coordinate $r$, and show that this yields black brane solutions carrying a new type of charge.
The field $A_t(r)$ is chosen to commute with the gauge field placed on $S^n$.
For example, when $n=4$, we can take an $SU(2)\times SU(2)$ gauge field on $S^4$, and choose a gauge field from any subgroup of $E_8$ that commutes with this $SU(2)\times SU(2)$. 
Thus the solution posesses electric and magnetic charges in different subsectors of the gauge group of the heterotic supergravity.
Remarkably, the black brane constructed here constitutes a rare example in which an exact solution can be obtained even with multiple gauge fields present — effectively overcoming the usual analytic intractability of such systems.
Moreover, for certain parameter values, the resulting spacetime exactly takes the form $\mathrm{AdS}_2\times S^n\times \mathbb{R}^{8-n}$, without the need to take a near-horizon limit, indicating the spontaneous emergence of an AdS geometry.

We also perform a perturbative analysis of the newly constructed black brane to study its dynamical instability.
Specifically, we introduce perturbations in the gauge sector and analyze the gauge field quasinormal modes (QNMs).
As an explicit example, we choose the time component $A_t$ within an $SU(2)$ gauge field, and show that as the temperature decreases, the original $U(1) \times U(1)$ symmetry is spontaneously broken to $U(1)$, signaling a phase transition.
The corresponding QNM represents an unstable mode, indicating the condensation of the $SU(2)$ gauge field and the transition to a new black brane solution.
This scenario is inspired by Gubser’s “colorful horizon” mechanism \cite{5}, and our result demonstrates that phase transitions can serve as a systematic tool to explore new black hole solutions in string theory.
Although we focus on the $SU(2)$ perturbations in this work, the constructed exact solution retains a sufficiently large flat sector $\mathbb{R}^{8-n}$, allowing us to consider larger gauge group components and analyze their stability as well.
If further instabilities are found, they will correspond to yet other black brane solutions — suggesting that our method provides a systematic approach to exploring the phase structure of complex black hole and black brane solutions.

Moreover, within the setup in which nontrivial gauge fields are placed on $S^n$, we propose a method for deriving further exact solutions and construct an analytic solution based on the approach. This is achieved by endowing part of the flat $\mathbb{R}^{8-n}$ directions with nontrivial geometry. Interestingly, the black brane solutions obtained in this way contain the geometry of a non-rotating BTZ black hole in three of the ten dimensions.

The structure of this paper is as follows.
In section \ref{section2}, we construct the new exact black brane solution.
As preparation, subsection \ref{subsec2.1} reviews the derivation of the previously known compactified (extremal) solutions \cite{3}.
In subsection \ref{subsec2.2}, we introduce an additional gauge field $A_t$ that commutes with the gauge field on $S^n$ and solve the corresponding equations of motion, obtaining new exact solutions for compactified $S^n$.
Section \ref{section3} analyzes the perturbative instability of these new solutions.
In subsection \ref{subsec3.1}, we review spontaneous symmetry breaking and phase transitions in Einstein-Yang-Mills black holes \cite{5}.
In subsection \ref{subsec3.2}, using the same gauge configurations, we perform a perturbation analysis and show that, as the temperature decreases, an unstable QNM appears, indicating gauge field condensation and a transition to a new black brane phase.
In Section \ref{sectionBTZ}, we introduce a strategy to obtain exact solutions in general situations and demonstrate its implementation by explicitly constructing a new analytic solution. 
Section \ref{section4} is devoted to a summary and discussions, with the black brane phase diagram which the new solution suggests. Appendix \ref{AppendixA} shows detailed calculations of the action and the equations of motion.

\section{Derivation of exact black brane solutions}\label{section2}
In this section, we construct exact black brane solutions. In subsection \ref{subsec2.1}, we review the derivation of the exact solution obtained in \cite{3}, and in subsection \ref{subsec2.2}, we derive a new exact solution by adding a temporal component of the gauge field, $A_t$. There, important features of the new solution are evaluated.

\subsection{Review of FKWY black brane solution}\label{subsec2.1}
Here, we review the derivation of the black brane solution constructed by Fukuda, Kobayashi, Watanabe, and Yonekura (FKWY) \cite{3}. Our new black brane solution is obtained by adding a temporal component of the gauge field, $A_t(r)$, which depends on the radial coordinate, to their solution.

In our setup, we decompose the $(n+1)$-dimensional spatial part of the ten-dimensional spacetime into the radial direction $r$ and an $n$-dimensional sphere $S^n$. Among the remaining $(9-n)$ dimensions, one is taken as the time direction $t$, while the remaining $p=8-n$ dimensions are kept flat. By placing a nontrivial gauge field configuration on $S^n$, we can construct a charged black brane solution. For example, when $n=4$, one can place an equal number of $SU(2)$ instantons and $SU(2)$ anti-instantons on $S^4$ in order to satisfy Bianchi identity : $\int_{S^n}dH=0$. So far, consistent gauge field configurations have been proposed only for the cases $n=1,2,4,8$. The Euclidean action in this setup is written as
\begin{align}
S=\int \dd[10]{x}\,\sqrt{G}\,e^{-2\Phi}
\left(
\mathcal{R}+4(\nabla\Phi)^2-\frac{\alpha'}{2}\operatorname{tr}|F|^2
\right),
\end{align}
where $G_{\mu\nu}$ is the metric, $\Phi$ is the dilaton, $\mathcal{R}$ is the scalar curvature, and $F_{\mu\nu}$ is the field strength of the gauge field.  
For the gauge field configuration mentioned above, if we denote the radius of $S^n$ by $R$, we have\footnote{See Appendix A.1 of \cite{3} for details of the gauge field configuration satisfying $\tr\abs{F}^2\propto R^{-4}$.} $\tr\abs{F}^2\propto R^{-4}$, and we can write
\begin{align}
\operatorname{tr}|F|^2=\frac{\mathsf{C}}{R^4},
\label{1}
\end{align}
where $\mathsf{C}>0$ is a constant\footnote{For example in the case of $n=4$, $\mathsf{C}$ is proportional to the instanton number of the gauge field.}.

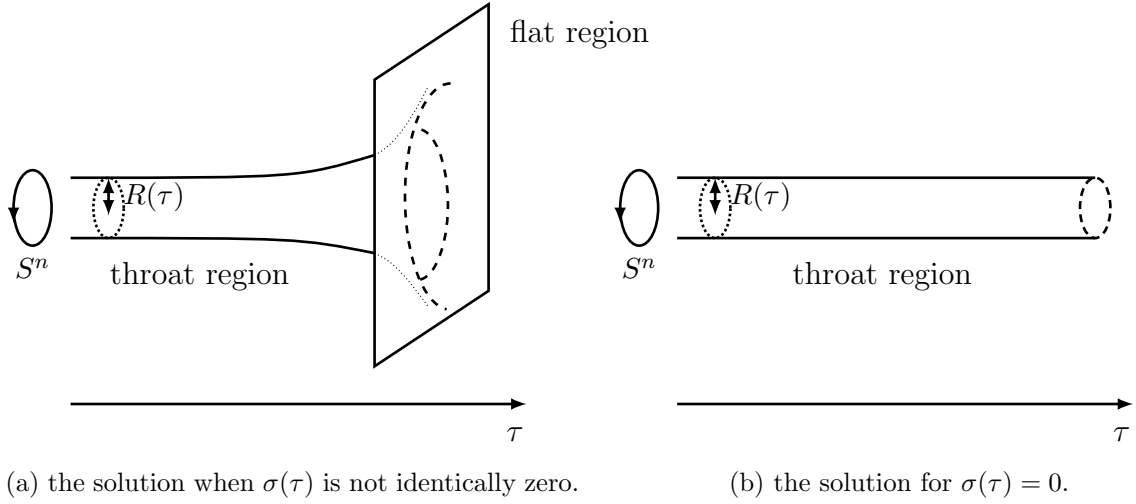
\begin{figure}
\begin{tabular}{cc}
\begin{minipage}[b]{0.5\columnwidth}
\begin{tikzpicture}[line width=1pt,>=latex]

\draw (-4,0.5) .. controls (-1,0.5) .. (0,0.8);
\draw[thin, densely dotted] (0,0.8)..controls (0.2,0.9) and (0.4,1.1)..(0.7,1.7);
\draw (-4,-0.3) .. controls (-1,-0.3).. (0,-0.5);
\draw[thin, densely dotted] (0,-0.5)..controls (0.2,-0.6) and (0.3,-0.65)..(0.7,-1.2);

\draw (0,-2) -- (0,1.8) -- (1.5,2.8) -- (1.5,-1) -- cycle;

\draw[dashed] (1,3.25) ++(0,-1.5)
  arc[start angle=90, end angle=265, x radius=0.6, y radius=1.5];
\draw[dashed] (0.6,0.65) ++(0,-1.5)
  arc[start angle=-85, end angle=85, x radius=0.4, y radius=1];

\draw[->] (-4,-2.5)--(2,-2.5);
\node at (1.85,-2.9) {\large{$\tau$}};

\node at (-2.3,-0.8) {\large{throat region}};
\node at (2.7,2.4) {\large{flat region}};

\draw (-4.5,0.1) ellipse (0.25 and 0.5);
\draw[->] (-4.75,0)--(-4.75,-0.1);
\node at (-4.5,-0.7) {$S^n$};

\draw[<->] (-3.5,0.5)--(-3.5,0);
\node at (-2.9,0.25) {$R(\tau)$};

\draw[densely dotted] (-3.5,0.1) ellipse (0.2 and 0.4);

\end{tikzpicture}
\subcaption{the solution when $\sigma(\tau)$ is not identically zero.}
\label{fig1a}
\end{minipage} & 
\begin{minipage}[b]{0.5\columnwidth}
\begin{tikzpicture}[line width=1pt,>=latex]]

\draw (-4,0.5)--(1.5,0.5);
\draw (-4,-0.3)--(1.5,-0.3);

\draw[densely dashed] (1.5,0.1) ellipse (0.2 and 0.4);

\draw[->] (-4,-2.5)--(2,-2.5);
\node at (1.85,-2.9) {\large{$\tau$}};

\node at (-1.3,-0.8) {\large{throat region}};

\draw (-4.5,0.1) ellipse (0.25 and 0.5);
\draw[->] (-4.75,0)--(-4.75,-0.1);
\node at (-4.5,-0.7) {$S^n$};

\draw[<->] (-3.5,0.5)--(-3.5,0);

\node at (-2.9,0.25) {$R(\tau)$};

\draw[densely dotted] (-3.5,0.1) ellipse (0.2 and 0.4);
\end{tikzpicture}
\subcaption{the solution for $\sigma(\tau)=0$.}
\label{fig1b}
\end{minipage}
\end{tabular}
\caption{Illustrations of the $\tau$-dependence of the radius $R$ of $S^n$.}
\label{fig1}
\end{figure}

Since we keep the $p=8-n$ dimensional space flat, a static and spherically symmetric metric can be written as
\begin{align}
\dd{s}^2=e^{2\Sigma(r)}\dd{t}_{\mathrm{E}}^2+N^2(r)\dd{r}^2+R^2(r)\dd{\Omega}_n^2+\dd{X}^i\dd{X}^i
\qquad (i=1,2,\cdots,p).
\end{align}
Introducing
\[
\ell_0\equiv\sqrt{\frac{\alpha'\mathsf{C}}{n(n-1)}},
\]
and defining the dimensionless coordinates
\begin{align}
\tilde{t}_{\mathrm{E}}\equiv\frac{t_{\mathrm{E}}}{\ell_0},\qquad
\tau\equiv\frac{r}{\ell_0\sqrt{\frac{8}{n(n-1)}}},
\end{align}
the metric becomes
\begin{align}
\dd{s}^2=\ell_0^2 e^{2\Sigma(\tau)}\dd{\tilde{t}}_{\mathrm{E}}^2
+\frac{8\ell_0^2}{n(n-1)}N^2(\tau)\dd{\tau}^2
+\ell_0^2 e^{2\sigma(\tau)}\dd{\Omega}_n^2+\dd{X}^i\dd{X}^i,
\label{2}
\end{align}
where $R=\ell_0 e^{\sigma(\tau)}$.  
Let $V_p$ denote the volume of the flat $p$-dimensional space.  
Substituting \eqref{1} and \eqref{2} into $S$, we obtain
\begin{align}
S&=V_p\int \dd{n+2}{x}\,\sqrt{G}\,e^{-2\Phi}
\left(
\mathcal{R}+4(\nabla\Phi)^2-\frac{\alpha'}{2}\operatorname{tr}|F|^2
\right)\nonumber\\
&\propto\int \dd{\tau}\,N\,e^{-2\varphi}
\left[
-\frac{n}{8}(N^{-1}\sigma')^2
-\frac{1}{8}(N^{-1}\Sigma')^2
+\frac{1}{2}(N^{-1}\varphi')^2
+e^{-2\sigma}-\frac{1}{2}e^{-4\sigma}
\right]\nonumber\\
&\equiv\int \dd{\tau}\,\mathcal{L},
\end{align}
where $\varphi\equiv\Phi-\frac{1}{2}\Sigma-\frac{n}{2}\sigma$, and the prime denotes differentiation with respect to $\tau$.  
From this expression, by taking the canonical form with respect to $\tau$ and fixing the gauge $N=1$, we obtain the following equations of motion\footnote{For detailed derivations, see Appendix A.}:
\begin{align}
&\frac{n}{4}(\sigma''-2\varphi'\sigma')-2(e^{-2\sigma}-e^{-4\sigma})=0, \label{3}\\
&\Sigma''-2\varphi'\Sigma'=0, \label{4}\\
&\varphi''-\varphi'^2-\frac{n}{4}\sigma'^2-\frac{1}{4}\Sigma'^2
+2e^{-2\sigma}-e^{-4\sigma}=0, \label{5}\\
&\frac{n}{4}\sigma'^2+\frac{1}{4}\Sigma'^2-\varphi'^2
+2e^{-2\sigma}-e^{-4\sigma}=0. \label{6}
\end{align}
\eqref{3}--\eqref{5} are the equations of motion for $\sigma$, $\Sigma$, and $\varphi$, while \eqref{6} represents the constraint associated with the gauge fixing $N=1$.  
The general solution of these equations was numerically obtained in \cite{3}, but here we focus on the special solution with $\sigma=0$.  
This solution, referred to as the “near-extremal, near-horizon limit” in \cite{3}, corresponds to a geometry where the throat region near the horizon extends infinitely, i.e., the radius $R$ of $S^n$ remains constant and does not grow\footnote{For the solution in which $\sigma(\tau)$ is not identically zero, it has been numerically confirmed in \cite{3} that $R$ exhibits a throat region and a flat region as shown in Figure \ref{fig1a}.}, see Figure \ref{fig1b}.
This solution with $\sigma=0$ can be obtained analytically.  
From \eqref{5} and \eqref{6}, we find
\begin{align}
\varphi''-2\varphi'^2+2=0,
\end{align}
whose solution is
\begin{align}
\varphi=-\frac{1}{2}\ln\sinh(2\tau)+\frac{1}{2}\ln c_1
\qquad (c_1:\mathrm{constant})
\label{7}
\end{align}
Here, $\tau=0$ corresponds to the horizon, where $e^{2\Sigma}=0$ is imposed as a boundary condition, and one integration constant is used to ensure $e^{-2\varphi(\tau=0)}=e^{-2\Phi(\tau=0)+\Sigma(\tau=0)}=0$.  
Rewriting \eqref{4} as $\frac{\dd{\Sigma'}}{\Sigma'}=2\varphi'\,\dd{\tau}$, we can integrate it to obtain
\begin{align}
&\Sigma'=D_1 e^{2\varphi}=\frac{D_1 c_1}{\sinh(2\tau)} \qquad (D_1:\mathrm{constant})\\
&\Sigma=\frac{D_1 c_1}{2}\ln\tanh\tau+\ln D_2 \qquad (D_2:\mathrm{constant})
\label{8}
\end{align}
and by requiring that \eqref{7} and \eqref{8} satisfy the constraint \eqref{6}, we find $D_1 c_1=2$.  
Thus, the near-horizon solution with $\sigma=0$ is
\begin{align}
\dd{s}^2&=\ell_0^2 D_2^2\tanh^2\tau\,\dd{\tilde{t}}_{\mathrm{E}}^2
+\frac{8\ell_0^2}{n(n-1)}\dd{\tau}^2
+\ell_0^2 \dd{\Omega}_n^2+\dd{X}^i\dd{X}^i \nonumber\\
&=D_2^2\tanh^2\mqty(\frac{r}{\ell_0\sqrt{\frac{8}{n(n-1)}}})
\dd{t}_{\mathrm{E}}^2+\dd{r}^2+\ell_0^2 \dd{\Omega}_n^2+\dd{X}^i\dd{X}^i.
\label{9}
\end{align}
By performing the coordinate transformation $t_{\mathrm{E}} \mapsto D_2 t_{\mathrm{E}}$ so that the spacetime approaches the normalized Euclidean space at infinity,  
$\dd{s}^2 = \dd{t_{\mathrm{E}}}^2 + \dd{r}^2 + \cdots$, and then redefining the time coordinate again as $t_{\mathrm{E}}$,  
we find that the metric becomes
\begin{align}
\dd{s}^2
= \tanh^2\!\mqty(\frac{r}{\ell_0\sqrt{\frac{8}{n(n-1)}}})\dd{t_{\mathrm{E}}}^2
+ \dd{r}^2
+ \ell_0^2\dd{\Omega_n^2}
+ \dd{X^i}\dd{X^i}. 
\label{corrected_metric_sol}
\end{align}
Near the horizon $r \simeq 0$, its behavior is
\begin{align}
\dd{s}^2 &\simeq
\mqty(\frac{r}{\ell_0\sqrt{\frac{8}{n(n-1)}}})^2
\dd{t_{\mathrm{E}}}^2
+ \dd{r}^2
+ \ell_0^2\dd{\Omega_n^2}
+ \dd{X^i}\dd{X^i}.
\end{align}
From the condition of no conical singularity at $r=0$, the temperature $T$ of the black brane is obtained as
\begin{align}
T = \sqrt{\frac{n(n-1)}{8}}\frac{1}{2\pi\ell_0}.
\label{11}
\end{align}
The dilaton $\Phi$ is given by
\begin{align}
\Phi = \varphi + \frac{1}{2}\Sigma
= -\ln\cosh\!\mqty(\frac{r}{\ell_0\sqrt{\frac{8}{n(n-1)}}})
+ \frac{1}{2}\ln\frac{D_2 c_1}{2}.
\label{10}
\end{align}
The obtained spacetime \eqref{9} and dilaton \eqref{10}, restricted to the $(t_{\mathrm{E}}, r)$ sector, coincide with the geometry known as Witten’s cigar geometry \cite{6}.  
A notable feature of this cigar geometry is that the constant part of the dilaton,  
$\Phi_0 = \frac{1}{2}\ln\frac{D_2 c_1}{2}$, which would normally have no physical effect,  
enters in the expression of the black hole mass $M$:
\begin{align}
M \propto e^{2\Phi_0} = \frac{D_2 c_1}{2}.
\end{align}

\subsection{New exact black brane solution}\label{subsec2.2}

We are ready to construct a solution by adding a gauge field $A_{\tilde{t}_\mathrm{E}}(\tau)\dd{\tilde{t}_\mathrm{E}}$ that commutes with the gauge field on $S^n$ to the black brane solution derived in subsection \ref{subsec2.1}.

\subsubsection{Equations and the solution}\label{subsubsec2.2.1}

We keep the same setup in which a gauge field is placed on an $n$-dimensional sphere while the $p$ spatial directions remain flat. The action is given by
\begin{align}
S=V_p\int\dd^{n+2}{x}\sqrt{G}e^{-2\Phi}\left(\mathcal{R}+4(\nabla\Phi)^2-\frac{\alpha^\prime}{2}\tr\abs{F}^2\right). \label{action}
\end{align}
Note that, since we introduce the new gauge field $A_{\tilde{t}_\mathrm{E}}(\tau)$, the term $\tr\abs{F}^2$ includes additional terms, thus it differs from that in subsection \ref{subsec2.1}.  
We choose the gauge $A_\tau=0$, and furthermore require that $A_{\tilde{t}_\mathrm{E}}$ commutes with the gauge field on $S^n$. For example, when $n=4$, we put $SU(2)\times SU(2)$ instantons on $S^4$, and hence we can choose a gauge group $H$ inside $E_8$ that commutes with this $SU(2)\times SU(2)$. 
The field $A_{\tilde{t}_\mathrm{E}}$ is then taken to be an element of the chosen group $H$. Consequently, the only nonvanishing components of the field strength $F_{\mu\nu}$, other than those on $S^n$ (denoted $F_{S^nS^{n\prime}}$), are
\begin{align}
F_{\tilde{t}_\mathrm{E}\tau}&=\partial_{\tilde{t}_\mathrm{E}}A_\tau-\partial_\tau A_{\tilde{t}_\mathrm{E}}-i[A_{\tilde{t}_\mathrm{E}},A_\tau] \nonumber\\
&=-A^\prime_{\tilde{t}_\mathrm{E}}.
\end{align}

Assuming again a static and spherically symmetric spacetime\footnote{Here $\ell_0$ and the dimensionless coordinates $\tilde{t}_\mathrm{E}$, $\tau$ are defined in the same way as in Section \ref{section2}.},
\begin{align}
\dd{s}^2&=e^{2\Sigma(r)}\dd{t_\mathrm{E}^2}+N^2(r)\dd{r^2}+R^2(r)\dd{\Omega_n^2}+\dd{X^i}\dd{X^i} \nonumber\\
&=\ell_0^2e^{2\Sigma(\tau)}\dd{\tilde{t}_\mathrm{E}^2}+\frac{8\ell_0^2}{n(n-1)}N^2(\tau)\dd{\tau^2}+\ell_0^2e^{2\sigma(\tau)}\dd{\Omega_n^2}+\dd{X^i}\dd{X^i},
\end{align}
we obtain\footnote{Strictly speaking, one should write $A_{\tilde{t}_\mathrm{E}}^a$ with an index for the gauge group basis, but we omit it below.}
\begin{align}
\tr\abs{F}^2&=\frac{\mathsf{C}}{R^4}+\mathrm{const}\cdot\tr(F_{\tilde{t}_\mathrm{E}\tau}F_{\tilde{t}_\mathrm{E}\tau}G^{\tilde{t}_\mathrm{E}\tilde{t}_\mathrm{E}}
G^{\tau\tau}) \nonumber\\
&=\frac{\mathsf{C}}{R^4}+\mathrm{const}\cdot A^{\prime 2}_{\tilde{t}_\mathrm{E}}e^{-2\Sigma}. \label{12}
\end{align}
For later convenience, we set the constant in \eqref{12} to be $\frac{1}{q^2}\frac{2}{\alpha^\prime}\frac{n(n-1)}{\ell_0^2}$, where $q$ is the gauge coupling constant.  
Taking the gauge $N=1$, the action can be written as
\begin{align}
S&\propto\int\dd{\tau}e^{-2\varphi}\left(-\frac{n}{8}\sigma^{\prime2}-\frac{1}{8}\Sigma^{\prime 2}+\frac{1}{2}\varphi^{\prime 2}+e^{-2\sigma}-\frac{1}{2}e^{-4\sigma}-\frac{1}{q^2}A^{\prime 2}_{\tilde{t}_\mathrm{E}}e^{-2\Sigma}\right) \\
&\equiv\int\dd{\tau}\mathcal{L}.
\end{align}
From the canonical form, the equations of motion and the constraint for $\sigma$, $\Sigma$, $\varphi$, and $A_{\tilde{t}_\mathrm{E}}$ are obtained as\footnote{See Appendix \ref{AppendixA} for detailed derivations.}
\begin{align}
&\frac{n}{4}(\sigma^{\prime\prime}-2\varphi^\prime\sigma^\prime)-2(e^{-2\sigma}-e^{-4\sigma})=0, \label{13}\\
&\Sigma^{\prime\prime}-2\varphi^\prime\Sigma^\prime=-\frac{8}{q^2}e^{-2\Sigma}A^{\prime 2}_{\tilde{t}_\mathrm{E}}, \label{14}\\
&\varphi^{\prime\prime}-\varphi^{\prime2}-\frac{n}{4}\sigma^{\prime 2}-\frac{1}{4}\Sigma^{\prime 2}+2e^{-2\sigma}-e^{-4\sigma}-\frac{2}{q^2}e^{-2\Sigma}A^{\prime 2}_{\tilde{t}_\mathrm{E}}=0, \label{15}\\
&A^{\prime\prime}_{\tilde{t}_\mathrm{E}}-2(\varphi^\prime+\Sigma^\prime)A^{\prime}_{\tilde{t}_\mathrm{E}}=0, \label{16}\\
&\frac{n}{4}\sigma^{\prime 2}+\frac{1}{4}\Sigma^{\prime 2}-\varphi^{\prime 2}+2e^{-2\sigma}-e^{-4\sigma}+\frac{2}{q^2}e^{-2\Sigma}A^{\prime 2}_{\tilde{t}_\mathrm{E}}=0. \label{17}
\end{align}
\eqref{13}--\eqref{16} are the equations of motion for each variable, while \eqref{17} represents the constraint.  

We again focus on the solution $\sigma=0$.  
From \eqref{15} and \eqref{17}, we obtain
\begin{align}
\varphi^{\prime\prime}-2{\varphi^{\prime}}^{2}+2=0,
\end{align}
and therefore the solution is found as
\begin{align}
\varphi=-\frac{1}{2}\ln\sinh(2\tau)+\frac{1}{2}\ln c_1. \label{18}
\end{align}
From \eqref{16}, we obtain
\begin{align}
A^{\prime}_{\tilde{t}_\mathrm{E}}=iCe^{2\varphi+2\Sigma}.  \label{19}
\end{align}
In the Euclidean theory, the temporal component of the gauge field $A_{\tilde{t}_\mathrm{E}}$ must be purely imaginary, and since $e^{2\varphi}$ and $e^{2\Sigma}$ are real, $C$ is an arbitrary real constant.  
Substituting \eqref{18} and \eqref{19} into \eqref{14} and \eqref{17} gives
\begin{align}
&\Sigma^{\prime\prime}+2\coth(2\tau)\Sigma^\prime=\frac{8C^2c_1^2}{q^2}\frac{e^{2\Sigma}}{\sinh^2(2\tau)}, \label{20}\\
&\Sigma^{\prime 2}-\frac{8C^2c_1^2}{q^2}\frac{e^{2\Sigma}}{\sinh^2(2\tau)}-\frac{4}{\sinh^2(2\tau)}=0. \label{21}
\end{align}
The function $\Sigma$ satisfying both of these equations is what we are seeking for.  
The constraint \eqref{21} can be rewritten as
\begin{align}
\Sigma^\prime=\pm\sqrt{1+\frac{2C^2c_1^2}{q^2}e^{2\Sigma}}\frac{2}{\sinh(2\tau)}, \label{22}
\end{align}
showing that $\Sigma(\tau)$ is either monotonically increasing or decreasing.  
We impose the initial condition that $\tau=0$ corresponds to the horizon, i.e., $e^{\Sigma(\tau=0)}=0$, or equivalently $\Sigma(\tau=0)=-\infty$.  
Thus, the branch with $\Sigma^\prime<0$ is unphysical, and we choose the positive sign in \eqref{22}.  
Then, rewriting
\begin{align}
\frac{\dd{\Sigma}}{\sqrt{1+\frac{2C^2c_1^2}{q^2}e^{2\Sigma}}}=\frac{2}{\sinh(2\tau)}\dd{\tau},
\end{align}
we can integrate to solve for $\Sigma$.  
The result, with an arbitrary constant $c_2$ ($0\leq c_2\leq 1$)\footnote{If $\abs{c_2}>1$, $G_{t_\mathrm{E}t_\mathrm{E}}$ diverges at finite $\tau$.}, is
\begin{align}
e^{2\Sigma}=\frac{2q^2c_2^2}{C^2c_1^2}\frac{\tanh^2\tau}{(1-c_2^2\tanh^2\tau)^2}.
\end{align}
Substituting this back into \eqref{20} confirms that it also satisfies the equation of motion, so this is the desired $\Sigma$.  
Integrating \eqref{19} with the expressions for $\varphi$ and $\Sigma$ gives
\begin{align}
A^{\prime}_{\tilde{t}_\mathrm{E}}&=i\frac{2q^2c_2^2}{Cc_1}\frac{\tanh^2\tau}{(1-c_2^2\tanh^2\tau)^2}\frac{1}{\sinh(2\tau)}, 
\end{align}
which is solved as
\begin{align}
A_{\tilde{t}_\mathrm{E}}&=-i\frac{q^2c_2^2}{2Cc_1}\frac{\sinh^2\tau}{c_2^2+(1-c_2^2)\cosh^2\tau}+\mathrm{const}. \label{24}
\end{align}
The dilaton $\Phi$ is given by
\begin{align}
\Phi&=\varphi+\frac{1}{2}\Sigma \\
&=-\frac{1}{2}\ln\sinh(2\tau)+\frac{1}{4}\ln\left(\frac{2q^2c_2^2}{C^2}\frac{\tanh^2\tau}{(1-c_2^2\tanh^2\tau)^2}\right). \label{25}
\end{align}
Its near-horizon behavior ($\tau\sim 0$) is $\Phi\sim -\frac{1}{2}\ln\tau+\frac{1}{4}\ln\tau^2+\mathrm{const}= \mathrm{const}$, showing that it remains finite there.  
For $\tau\to\infty$ and $c_2\neq 1$, we find $\Phi\sim -\tau$, indicating that the coupling becomes weak at large distance. The special case $c_2=1$ will be studied separately in section \ref{subsubsec2.2.2}.

Here again, we perform the coordinate transformation  
$t_{\mathrm{E}} \mapsto \frac{\sqrt{2}\, q c_2}{C c_1 (1-c_2^2)}\, t_{\mathrm{E}}$  
so that the spacetime becomes the normalized Euclidean space at infinity,  
$\dd{s}^2 = \dd{t_{\mathrm{E}}}^2 + \dd{r}^2 + \cdots$.  
Writing the new time coordinate again as $t_{\mathrm{E}}$ (or in the dimensionless form  
$\tilde{t}_{\mathrm{E}} = t_{\mathrm{E}}/\ell_0$), we find that the metric becomes
\begin{align}
\dd{s}^2 &=
\frac{\ell_0^2 (1-c_2^2)^2 \tanh^2 \tau}{(1-c_2^2 \tanh^2 \tau)^2}
\,\dd{\tilde{t}_{\mathrm{E}}}^2
+ \frac{8\ell_0^2}{n(n-1)}\dd{\tau^2}
+ \ell_0^2 \dd{\Omega_n^2}
+ \dd{X^i}\dd{X^i},
\label{metric_sol}
\end{align}
and the gauge field is
\begin{align}
A_{\tilde{t}_{\mathrm{E}}}
= -i\,\frac{q c_2 (1-c_2^2)}{2\sqrt{2}}
\,\frac{\sinh^2 \tau}{c_2^2 + (1-c_2^2)\cosh^2\tau}
+ \mathrm{const}.
\label{gauge_sol}
\end{align}
Consequently, the metric, the gauge field, and the dilaton can be collectively written as follows:
\begin{tcolorbox}
\begin{align}
\dd{s}^2 &=
\frac{\ell_0^2 (1-c_2^2)^2 \tanh^2 \tau}{(1-c_2^2 \tanh^2 \tau)^2}
\,\dd{\tilde{t}_{\mathrm{E}}}^2
+ \frac{8\ell_0^2}{n(n-1)}\dd{\tau^2}
+ \ell_0^2 \dd{\Omega_n^2}
+ \dd{X^i}\dd{X^i}, 
\nonumber\\
A_{\tilde{t}_{\mathrm{E}}}
&= -i\,\frac{q c_2 (1-c_2^2)}{2\sqrt{2}}
\,\frac{\sinh^2 \tau}{c_2^2 + (1-c_2^2)\cosh^2\tau}
+ \mathrm{const}, 
\label{exact_solution}\\
\Phi&=-\frac{1}{2}\ln\sinh(2\tau)+\frac{1}{4}\ln\left(\frac{2q^2c_2^2}{C^2}\frac{\tanh^2\tau}{(1-c_2^2\tanh^2\tau)^2}\right).
\nonumber
\end{align}
\end{tcolorbox}
\noindent
This is our new exact black brane solution.

\subsubsection{Charge, mass and temperature}\label{subsubsec2.2.2}

It is rather important to note that a solution of exactly the same form appears in two-dimensional heterotic string theory. We make use of this for extracting the physical quantities of the new solution.

The action of two-dimensional heterotic supergravity, written with the Lorentzian signature, is
\begin{align}
S = \int \dd^2 x \sqrt{-G}\, e^{-2\Phi}\!
\left(
\mathcal{R} - c + (\nabla \Phi)^2 - \tr|F|^2
\right),
\end{align}
where $c$ $(<0)$ is the central charge resulting from the target space not being at the critical dimension, and since we are in two dimensions, there is no antisymmetric $B$-field contribution.  
The solution for the case of a $U(1)$ gauge field was obtained in \cite{McGuigan:1991qp, Gibbons:1992rh}, and in terms of free parameters $\mathsf{m}, \mathsf{q} (\mathsf{m}>\mathsf{q}), \Phi_0$, it is given by
\begin{align}
&\dd{s}^2
= -\frac{(\mathsf{m}^2-\mathsf{q}^2)\sinh^2 2\lambda r}
{(\mathsf{m}+\sqrt{\mathsf{m}^2-\mathsf{q}^2}\cosh^2 2\lambda r)^2}\dd{t}^2
+ \dd{r}^2,
\label{2dim_metric}\\[6pt]
&A_t
= \frac{\sqrt{2}\,\mathsf{q}}
{(\mathsf{m}+\sqrt{\mathsf{m}^2-\mathsf{q}^2}\cosh^2\lambda r)^2}
- \frac{\sqrt{2}\,\mathsf{q}}{(\mathsf{m}+\sqrt{\mathsf{m}^2-\mathsf{q}^2})^2},
\label{2dim_gauge}\\[6pt]
&\Phi
= \Phi_0
- \frac{1}{2}
\ln\!\left[
\frac{1}{2}
\left(
\frac{\mathsf{m}}{\sqrt{\mathsf{m}^2-\mathsf{q}^2}}
+ \cosh 2\lambda r
\right)
\right],
\label{2dim_dilaton}
\end{align}
where $c = -\frac{\lambda^2}{4}$\footnote{The geometric properties of this spacetime are discussed in detail in
~\cite{Giveon:2004zz, Perry:1993ry, Yi:1993gh}.
}. 
By taking   
$c_2^2 = \frac{\mathsf{m}-\sqrt{\mathsf{m}^2-\mathsf{q}^2}}
{\mathsf{m}+\sqrt{\mathsf{m}^2-\mathsf{q}^2}}$,  
$\lambda = \sqrt{\frac{n(n-1)}{8\ell_0^2}}$,  
and $\Phi_0 = \frac{1}{2}\ln\frac{q c_2}{\sqrt{2} C (1-c_2^2)}$,  
and then appropriately rewriting the expressions, the $(t,r)$ part of our solution \eqref{exact_solution} exactly agrees with theirs\footnote{
Although there appear to be two parameters $\mathsf{m}$ and $\mathsf{q}$, only the ratio $\mathsf{q}/\mathsf{m}$ enters physical quantities.  
Also note that the notation in \cite{McGuigan:1991qp, Gibbons:1992rh} differs from that used in this paper, in particular concerning the prefactors of the dilaton.  
Equations \eqref{2dim_metric}, \eqref{2dim_gauge}, \eqref{2dim_dilaton} are written after adjusting their notation to ours.
}.

This correspondence arises because, in order to obtain an exact solution, we focused on the case $\sigma(\tau)=0$, namely the solution with constant radius $R=\mathrm{const}$ of the $S^n$.  
Since $R$ is constant, the contribution $-\tr|F|^2$ from the gauge field on $S^n$ becomes a constant.  
Further, the total ten-dimensional curvature decomposes as  
$\mathcal{R} = \mathcal{R}_{(t,r)} + \mathcal{R}_{S^n}$,  
and the $S^n$ curvature is also constant,  
$\mathcal{R}_{S^n} = \frac{n(n-1)}{R^2}$.  
The sum of these acts in the same way as the central charge $c$ in the two-dimensional heterotic supergravity.  
Thus, upon reduction of our setup to two dimensions in the $(t,r)$ sector, the effective action takes exactly the same form as heterotic supergravity in the anomalous dimension $D=1+1$.  
Consequently, the resulting solutions must coincide, and therefore the charge and mass of our black brane may be computed using the results of \cite{McGuigan:1991qp, Gibbons:1992rh}.

\paragraph{\underline{Charge}}
\quad 
Let us compute the charge $Q$ of our ten-dimensional black brane solution. 
Since the $\mathbb{R}^p$ directions parallel to the brane are flat, the physically relevant quantity is the charge density per unit brane volume, defined as 
\begin{align}
Q_{n+2} \equiv Q_{10}/{V_p},
\end{align}
where $V_p$ is the volume of the $\mathbb{R}^p$ directions. 
The quantity $Q_{n+2}$ is obtained by integrating the gauge flux over $S^n$ evaluated at $\tau \to \infty$. 
In the present solution, the radius of $S^n$ is a constant $\ell_0$, and the gauge field is spherically symmetric. 
Therefore, the flux integral over $S^n$ simply reduces to multiplying the asymptotic flux value $Q_2$ by $A_n$, which is the surface area of the $n$-sphere of radius $\ell_0$, namely
\begin{align}
Q_{n+2} = A_n Q_2 .
\end{align}
Here, $Q$ also represents the charge of the two-dimensional $(t, r)$ sector, which was computed in \cite{Gibbons:1992rh} and is given by

\begin{align}
Q_2= \frac{\sqrt{n(n-1)}}{\ell_0}\,
\frac{c_2}{1-c_2^2}
\,e^{-2\Phi_0}, 
\end{align}
where we have used the relation  
$\frac{\mathsf{q}}{\mathsf{m}}=\frac{2c_2}{1+c_2^2}$ in the intermediate steps.  
From now on, since $C$ is an arbitrary constant, we regard $\Phi_0$ as an arbitrary constant independent of $c_2$.\\
In summary, the ten-dimensional charge $Q$ is given by
\begin{align}
Q= A_nV_p\frac{\sqrt{n(n-1)}}{\ell_0}\,
\frac{c_2}{1-c_2^2}
\,e^{-2\Phi_0}.
\label{charge}
\end{align}

\paragraph{\underline{Mass}}
\quad
As in the case of the charge, the mass of the ten-dimensional black brane can also be reduced to the ADM mass of the two-dimensional $({t}, r)$ sector. 
Let us define the mass density by dividing the ten-dimensional mass $M$ by the volume $V_p$ of the $\mathbb{R}^p$ directions,
$M_{n+2} \equiv M/V_p$.
The quantity $M_{n+2}$ can be defined using the standard canonical formalism in $(n+2)$ dimensions. 
We choose the reference metric for the energy computation to be
\begin{align}
ds^2 = dt_\mathrm{E}^2 + dr^2 + \ell_0^2 d\Omega_n^2 .
\end{align}
The energy of the spacetime is obtained from an integral over $S^n$ evaluated on the boundary $\tau \to \infty$. 
Since the radius of $S^n$ is constant and equal to $\ell_0$, this integral simply reduces to the surface area $A_n$ of $S^n$ multiplied by the energy density $M_2$ of the two-dimensional $(t, r)$ sector, namely
\begin{align}
M_{n+2} = A_n M_2 .
\end{align}
Here $M_2$ is the ADM mass of the two-dimensional $(t, r)$ sector, which was computed in \cite{Gibbons:1992rh} and is given by
\begin{align}
M_2 = \sqrt{\frac{n(n-1)}{2\ell_0^2}}\,\frac{1+c_2^2}{1-c_2^2}\,e^{-2\Phi_0}.
\end{align}
Putting everything together, we obtain
\begin{align}
M = A_n V_p\sqrt{\frac{n(n-1)}{2\ell_0^2}}\,\frac{1+c_2^2}{1-c_2^2}\,e^{-2\Phi_0} .
\label{mass}
\end{align}

As an important caveat, since the dilaton $\Phi$ diverges to $-\infty$ as $\tau \to \infty$, a naive computation of the ADM mass of the $(t, r)$ sector would lead to a divergent result. 
This divergence can be cured by introducing an appropriate boundary counterterm in the action that cancels the dilaton contribution. 
More explicitly, this two-dimensional system admits not only the black hole solution but also a two-dimensional Minkowski spacetime (or a two-dimensional Euclidean space after Euclideanization) with a linear dilaton profile as a solution. 
The counterterm is chosen such that the mass of this Minkowski background is normalized to zero, and the mass of the black hole solution is then defined relative to this reference background. 
For a more detailed discussion, see \cite{Gibbons:1992rh}.

\paragraph{\underline{Temperature}}
Near $\tau \simeq 0$, the black brane metric behaves as
\begin{align}
\dd{s}^2 &=
\ell_0^2\frac{(1-c_2^2)^2 \tanh^2\tau}{(1-c_2^2\tanh^2\tau)^2}
\dd{\tilde{t}_{\mathrm{E}}}^2
+ \frac{8\ell_0^2}{n(n-1)}\dd{\tau^2}
+ \ell_0^2\dd{\Omega_n^2}
+ \dd{X^i}\dd{X^i}
\label{23}\\
&\simeq
(1-c_2^2)^2
\mqty(\frac{r}{\ell_0\sqrt{\frac{8}{n(n-1)}}})^2\dd{t_\mathrm{E}}^2
+ \dd{r}^2
+ \ell_0^2\dd{\Omega_n^2}
+ \dd{X^i}\dd{X^i},
\end{align}
and therefore the temperature $T$ of the black brane is
\begin{align}
T=\sqrt{\frac{n(n-1)}{8}}\,\frac{1-c_2^2}{2\pi \ell_0}.
\label{temp}
\end{align}

\subsubsection{Limits of the new solution}\label{subsubsec2.2.3}

We study two interesting limits of the parameter $c_2$ in the solution; one is the no charge limit with $c_2=0$, and the other is the extremal limit with $c_2=1$ which is the exact $\mathrm{AdS}_2$ limit. These two limits will be used in section \ref{section3} for our perturbation analyses.

\paragraph{\underline{No charge limit}}
It is natural that the solution obtained in subsection \ref{subsec2.2} reduces continuously to the chargeless FKWY solution reviewed in subsection \ref{subsec2.1} when the charge is taken to zero.  
Since the charge $Q$ vanishes at $c_2=0$, one may say, roughly speaking, that increasing $c_2$ corresponds to increasing the charge\footnote{
However, as is clear from the expression for $M$, in this system one cannot vary the charge and mass independently.
}.
Indeed, by setting $c_2=0$, one confirms that the metric \eqref{metric_sol} reduces to the uncharged solution \eqref{corrected_metric_sol}.  
Moreover, for $c_2=0$, the gauge field \eqref{gauge_sol} becomes $0+\mathrm{const}$.  
For the dilaton, from
\begin{align}
\Phi = \Phi_0 - \frac{1}{2}\ln\left[
\frac{1}{2}
\left(
\frac{1+c_2^2}{1-c_2^2}
+ \cosh 2\tau
\right)
\right],
\end{align}
one sees that at $c_2=0$ it matches the form of \eqref{10}.

\paragraph{\underline{Extremal (exact $\mathrm{AdS}_2$) limit}}
The temperature $T$ in \eqref{temp} becomes zero when $c_2=1$, but in this limit the behavior of the metric \eqref{metric_sol} becomes ill-defined.  
This can be resolved by the coordinate transformation  
$\tilde{t}_{\mathrm{E}} \mapsto (1-c_2^2)\tilde{t}_{\mathrm{E}}$.  
Writing the transformed time coordinate again as $\tilde{t}_{\mathrm{E}}$, the metric \eqref{23} at $c_2=1$ is calculated as 
\begin{align}
\dd{s}^2 &=
\ell_0^2\sinh^2(2\tau)\dd{\tilde{t}_{\mathrm{E}}}^2
+ \frac{8\ell_0^2}{n(n-1)}\dd{\tau^2}
+ \ell_0^2\dd{\Omega_n^2}
+ \dd{X^i}\dd{X^i} \nonumber \\[4pt]
&=
\frac{2\ell_0^2}{n(n-1)}
\left[
\frac{n(n-1)}{2}\sinh^2\tilde{\tau}\,
\dd{\tilde{t}_{\mathrm{E}}}^2
+ \dd{\tilde{\tau}}^2
\right]
+ \ell_0^2\dd{\Omega_n^2}
+ \dd{X^i}\dd{X^i},
\qquad (\tilde{\tau}\equiv 2\tau)
\label{40}
\end{align}
which, after restoring Lorentzian signature, corresponds to  
$\mathrm{AdS}_2 \times S^n \times \mathbb{R}^p$.  
Here, the $\mathrm{AdS}_2$ is written in the black-hole patch.

The dilaton in this limit is obtained by inserting $c_2=1$ into \eqref{25}, yielding
\begin{align}
\Phi = \frac{1}{4}\ln\frac{q^2}{2C^2},
\end{align}
which is constant.  
For the gauge field, applying the coordinate transformation  
$\tilde{t}_{\mathrm{E}} \mapsto (1-c_2^2)\tilde{t}_{\mathrm{E}}$ to \eqref{24} and then setting $c_2=1$, we obtain
\begin{align}
A_{\tilde{t}_{\mathrm{E}}}
= -i\,\frac{q}{2\sqrt{2}}\sinh^2\tau
+ \mathrm{const}.
\label{AdS2_gauge}
\end{align}

\section{Instability and phase transition of the black brane}\label{section3}

In this section, we perform the stability analysis on the new black brane solutions with $n=2,4$ constructed in section \ref{section2}.  
In subsection \ref{subsec3.1}, we review the spontaneous symmetry breaking and phase transition in black holes of the Einstein--Yang--Mills system~\cite{5}, which is one of the motivations for our instability analysis.  
In subsection \ref{subsec3.2}, by introducing a perturbation of the gauge field configuration similar to those used in~\cite{5}, we show that an unstable QNM (quasinormal mode) appears.  
This instability implies that the perturbed gauge field grows macroscopically, leading to a phase transition to a new black brane configuration.

\subsection{Review of the phase transition of RN--AdS black hole}\label{subsec3.1}

Within the framework of gravity theories with a negative cosmological constant, asymptotically AdS spacetimes have been extensively studied from the viewpoint of holography.  
Among them, the Reissner--Nordstr\"{o}m (RN) AdS black hole, which is a solution of the Einstein--Maxwell theory with a negative cosmological constant, plays a crucial role in the framework of the so-called “holographic superconductor” \cite{4, Gubser:2008px, Hartnoll:2008vx, Albash:2008eh, 5, Gubser:2008wv, Maldacena:1997re, Witten:1998qj}, which aims to understand superconductivity holographically. 
The electromagnetic $U(1)$ symmetry associated with electrons is spontaneously broken when two electrons form a Cooper pair and condense, and an attempt to reproduce this phenomenon holographically was first made in~\cite{Gubser:2008px}.  
When a scalar field $\psi$ with mass $m$ and charge $q$ is introduced perturbatively in the RN $\mathrm{AdS}_4$ black hole background, the Lagrangian $\mathcal{L}$ takes the form
\begin{align}
\mathcal{L}
=R-\frac{6}{L^2}-\frac{1}{4}F_{\mu\nu}^2-\abs{\partial_\mu\psi - iqA_\mu{\psi}}^2 - m^2\abs{\psi}^2,
\end{align}
and the effective mass $m_{\mathrm{eff}}$ of the scalar field $\psi$ is given by
\begin{align}
m_{\mathrm{eff}}^2 = m^2 + g^{tt}q^2{A_0^2}.
\end{align}
Here, $R$ is the Ricci scalar, $L$ is the AdS radius, and $A_\mu$ denotes the background gauge field.  
In~\cite{Gubser:2008px}, it was argued that $m_{\mathrm{eff}}^2$ falls below the BF bound~\cite{Breitenlohner:1982jf}$\,$ near the horizon, causing the scalar field to become unstable and condense.  
In the boundary gauge theory, this condensation corresponds to the superconducting phase transition, which is interpreted as the $s$-wave holographic superconductor scenario.

While $s$-wave superconductivity is understood as the condensation of a scalar field, $p$-wave superconductivity is interpreted as the condensation of a vector field.  
The first holographic model for $p$-wave superconductivity was proposed in~\cite{5}, where the RN--AdS black hole was extended to include a non-Abelian gauge field, leading to an asymptotically AdS black hole with $SU(2)$ gauge symmetry.  
The Lagrangian is given by
\begin{align}
\mathcal{L}=R-\frac{6}{L^2}-\frac{1}{4}(F_{\mu\nu}^a)^2, \label{3.1.1}
\end{align}
where the gauge group is taken to be $SU(2)$.  
Using the $SU(2)$ generators $\lambda^a = \sigma^a / 2i$ ($\sigma^a$ being the Pauli matrices), we have
\begin{align}
&[\lambda^a, \lambda^b] = \epsilon^{abc}\lambda^c, \\
&F_{\mu\nu}^a = \partial_\mu A_\nu^a - \partial_\nu A_\mu^a + g\,\epsilon^{abc}A_\mu^b A_\nu^c.
\end{align}
We then seek for solutions of the equations of motion derived from \eqref{3.1.1} in the following form.  
The metric ansatz is
\begin{align}
\dd{s}^2 = e^{2a(r)}\left(-h(r)\dd{t}^2 + (\dd{x^1})^2 + (\dd{x^2})^2\right) + \frac{\dd{r}^2}{e^{2a(r)}h(r)},
\end{align}
and the gauge field $A = A_\mu^a\lambda^a\dd{x^\mu}$ is taken as
\begin{align}
A = \phi(r)\lambda^3\dd{t} + w(r)\left(\lambda^1\dd{x^1} + \lambda^2\dd{x^2}\right). \label{3.1.2}
\end{align}
Since $\lambda^3$ generates the $U(1)$ subgroup of $SU(2)$, the solution with $w=0$ corresponds to the RN $\mathrm{AdS}_4$ black hole with a $U(1)$ charge :
\begin{align}
\dd{s}^2
= -\left(1-\frac{2M}{r} +\frac{Q^2}{r^2} +\frac{r^2}{L^2}\right)\,\dd{t}^2 + \frac{\dd{r}^2}{\left(1-\frac{2M}{r} +\frac{Q^2}{r^2} +\frac{r^2}{L^2}\right)} + r^2 \dd{\Omega_{2}^2}.
\end{align}
Under a gauge transformation generated by this $U(1)$ subgroup, the components $\lambda^1$ and $\lambda^2$ transform as
\begin{align}
\mqty(\lambda^1 \\ \lambda^2)
\mapsto
\mqty(\cos\theta & -\sin\theta \\ \sin\theta & \cos\theta)
\mqty(\lambda^1 \\ \lambda^2).
\end{align}
For $w=0$, the gauge field $A$ remains invariant under this $U(1)$ transformation, and both the metric and $A$ are invariant under the spatial rotation $SO(2)\cong U(1)$ in the $(x^1, x^2)$ plane.  
Thus, the solution possesses a $U(1)\times U(1)$ symmetry.

On the other hand, if a solution with $w\neq0$ exists, the gauge field $A$ is no longer invariant under the $U(1)\times U(1)$ transformations.  
However, under a simultaneous rotation by the same angle $\theta$ in both the internal and spatial $(x^1, x^2)$ spaces,
\begin{align}
\mqty(\lambda^1 \\ \lambda^2)
\mapsto
\mqty(\cos\theta & -\sin\theta \\ \sin\theta & \cos\theta)
\mqty(\lambda^1 \\ \lambda^2), \qquad
\mqty(x^1 \\ x^2)
\mapsto
\mqty(\cos\theta & -\sin\theta \\ \sin\theta & \cos\theta)
\mqty(x^1 \\ x^2),
\end{align}
the gauge field $A$ becomes invariant.  
This means that the symmetry is spontaneously broken from $U(1)\times U(1)$ down to $U(1)$.  
In~\cite{5}, the equations of motion for $a(r)$, $h(r)$, $\phi(r)$, and $w(r)$ were solved numerically, and a solution with a nonzero $A$ condensate near the horizon was obtained.  
This can be interpreted as a condensation of the gauge field due to spontaneous symmetry breaking.  
Since the non-Abelian gauge field develops its nonvanishing profile near the horizon, this phenomenon was referred to as a “colorful horizon” in~\cite{5}.  
The holographic dual of this black hole solution is interpreted as the $p$-wave superconductor.

\subsection{Phase transition of the new black brane}\label{subsec3.2}

The scenario reviewed in subsection \ref{subsec3.1} can be applied not only to holographic superconductors but also to the analysis of pure gravitational dynamics. Namely, we consider the planar part of a black hole possessing only the $A_t$ component, and introduce a perturbation $w$ of the form \eqref{3.1.2}. If there exists an unstable mode for this $w$, it implies that the black hole is dynamically unstable and the system favors a new configuration where the gauge field condenses spontaneously, breaking the symmetry. In the following, we apply this idea to the new black brane solution we constructed in subsection \ref{subsec2.2}.

\subsubsection{Perturbation equation}\label{subsubsec3.2.1}

We choose $SU(2)$ as the gauge group $H$ to which $A_{\tilde{t}_\mathrm{E}}$ belongs. For the solution obtained in subsection \ref{subsec2.2}, we consider perturbations of the gauge field in the $X^1$ and $X^2$ directions of the flat part $\dd{s}^2=\cdots+\dd{X^i}\dd{X^i}$. For convenience, we introduce a dimensionless coordinate $x^i\equiv X^i/\ell_0$ and consider the Lorentzian black brane
\begin{align}
\dd{s}^2=-\frac{(1-c_2^2)^2\tanh^2\tau}{(1-c_2^2\tanh^2\tau)^2}\dd{\tilde{t}}^2+\frac{8\ell_0^2}{n(n-1)}\dd{\tau^2}+\ell_0^2\dd{\Omega_n^2}+\ell_0^2(\dd{x^1})^2&+\ell_0^2(\dd{x^2})^2+\dd{X^i}\dd{X^i},\\
&(i=3,4,\cdots, p) \notag
\end{align}
with the gauge field
\begin{align}
A=\phi(\tau)\lambda^3\dd{\tilde{t}}+w(\tilde{t}, \tau)(\lambda^1\dd{x^1}+\lambda^2\dd{x^2}), \label{26}
\end{align}
where $\phi(\tau)\lambda^3$ is the gauge field derived in subsection \ref{subsec2.2},
\begin{align}
\phi(\tau)=\frac{qc_2(1-c_2^2)}{2\sqrt{2}}\frac{\sinh^2\tau}{c_2^2+(1-c_2^2)\cosh^2\tau},
\end{align}
which takes values in $U(1)\subset SU(2)$. The term $w(\tau)\lambda^i$ $(i=1,2)$ corresponds to a perturbation added on top of the background black brane solution constructed in subsection \ref{subsec2.2}.

In the notation $A=A^a_\mu\lambda^a\dd{x}^\mu$, the nonzero components are
\begin{align}
A_{\tilde{t}}^3=\phi(\tau),\quad A_{x^1}^1=w(\tilde{t}, \tau),\quad A_{x^2}^2=w(\tilde{t}, \tau),
\end{align}
and the nonvanishing components of
$F_{\mu\nu}=F_{\mu\nu}^a\lambda^a=(\partial_\mu A_\nu^a-\partial_\nu A_\mu^a+\epsilon^{abc}A_\mu^b A_\nu^c)\lambda^a$
are
\begin{align}
&F_{\tau x^1}^{1}=w',\quad F_{{\tilde{t}} x^2}^{1}=-\phi w,\quad F_{{\tilde{t}} x^1}^{1}=\partial_{\tilde{t}}w,\nonumber\\
&F_{\tau x^2}^{2}=w',\quad F_{{\tilde{t}} x^1}^{2}=\phi w,\quad F_{{\tilde{t}} x^2}^{2}=\partial_{\tilde{t}}w,\nonumber\\
&F_{\tau \tilde{t}}^{3}=\phi',\quad F_{x^1 x^2}^{3}=w^2.
\end{align}

With the evaluation
\begin{align}
F_{\mu\nu}^a F_{\rho\sigma}^a G^{\mu\rho}G^{\nu\sigma}=\frac{2}{\ell_0^4}\left[\frac{n(n-1)}{4}w'^{2}-2\phi^2w^2e^{-2\Sigma}-\frac{n(n-1)}{8}\phi'^{2}e^{-2\Sigma}-2(\partial_{\tilde{t}}w)^2e^{-2\Sigma}+w^4\right],
\end{align}
the Lagrangian $\mathcal{L}_{SU(2)}$ of the $SU(2)$ gauge field becomes
\begin{align}
\mathcal{L}_{SU(2)}&\propto\sqrt{-G}e^{-2\Phi}\left[\frac{n(n-1)}{4}w'^{2}-2\phi^2w^2e^{-2\Sigma}-\frac{n(n-1)}{8}\phi'^{2}e^{-2\Sigma}-2(\partial_{\tilde{t}}w)^2e^{-2\Sigma}+w^4\right]\nonumber\\
&\propto e^{-2\varphi}\left[\frac{n(n-1)}{4}w'^{2}-2\phi^2w^2e^{-2\Sigma}-\frac{n(n-1)}{8}\phi'^{2}e^{-2\Sigma}-2(\partial_{\tilde{t}}w)^2e^{-2\Sigma}+w^4\right].
\end{align}
Here we are using the shorthand notation for the components of the metric,
$\dd{s}^2 = -\ell_0^2 e^{2\Sigma} \dd{\tilde{t}}^2 + \frac{8\ell_0^2}{n(n-1)} \dd{\tau}^2 + \ell_0^2 \dd{\Omega_n}^2 + \ell_0^2(\dd{x^1})^2 + \ell_0^2(\dd{x^2})^2 + \cdots$.

From the Euler--Lagrange equation, neglecting $\mathcal{O}(w^2)$ terms, we obtain
\begin{align}
\frac{n(n-1)}{8}e^{2\Sigma}w''-\frac{n(n-1)}{4}\varphi'e^{2\Sigma}w'+\phi^2w=(\partial_{\tilde{t}}^2w). \label{28}
\end{align}
Separating variables as $w(\tilde{t},\tau)=f(\tau)g(\tilde{t})$, we get $\partial_{\tilde{t}}^2 g = -\omega^2 g$ with the solution $g(\tilde{t}) = e^{\pm i\omega\tilde{t}}$. Substituting $w=f(\tau)e^{-i\omega\tilde{t}}$ and the expressions for $\varphi, \phi, e^{2\Sigma}$ into \eqref{28}, we obtain
\begin{align}
f^{\prime\prime}(\tau)+2\coth(2\tau)f^\prime(\tau)+\frac{q^2c_2^2}{n(n-1)}\sinh^2\tau\cosh^2\tau\left[\frac{1-c_2^2\tanh^2\tau}{c_2^2+(1-c_2^2)\cosh^2\tau}\right]^2f(\tau)& \notag\\
=-\frac{8\omega^2}{n(n-1)}\frac{(1-c_2^2\tanh^2\tau)^2}{(1-c_2^2)^2\tanh^2\tau}f(\tau),& \label{27}
\end{align}
which forms an eigenvalue equation for $\omega$. If a solution $(f, \omega)$ with $\Im(\omega) > 0$ exists for this equation, then it represents an unstable mode, which means that the black brane spontaneously breaks the symmetry. 

We also expect a phase transition: we lower the temperature $T=\sqrt{\frac{n(n-1)}{8}}\frac{1-c_2^2}{2\pi\ell_0}$ and examine whether the system moves from a stable phase to an unstable phase during this process. In other words, we would like to investigate whether an unstable mode appears when $c_2$ is varied over $[0,1]$, to see the phase transition. In general it is difficult to analytically solve and analyze \eqref{27}, and numerically searching for the value of $c_2$ at which an unstable mode emerges is also laborious. Therefore, we focus here on whether a phase transition exists and examine the stability at the endpoints $c_2=0$ and $c_2=1$ separately. If the system is stable at $c_2=0$ and an unstable mode exists at $c_2=1$, then a phase transition must occur somewhere in between.

\subsubsection{Quasinormal mode and its instability}\label{subsubsec3.2.2}

\paragraph{\underline{The case $c_2=0$}}

This corresponds to the no charge limit of the solution in section \ref{subsubsec2.2.2}. Equation \eqref{27} reduces to
\begin{align}
f^{\prime\prime}(\tau)+2\coth(2\tau)f^\prime(\tau)=-\frac{8\omega^2}{n(n-1)}\coth^2\tau f(\tau). \label{29} 
\end{align}
With the tortoise coordinate $\tau_*$ defined via $\dv{\tau_*}{\tau}=\sqrt{\frac{8}{n(n-1)}}e^{-\Sigma}=\sqrt{\frac{8}{n(n-1)}}\coth\tau$, \eqref{29} becomes
\begin{align}
\dv[2]{f}{\tau_*}+\left[\sqrt{\frac{n(n-1)}{2}}\tanh^2\tau\right]\dv{f}{\tau_*}=-\omega^2 f. \label{46}
\end{align}
Letting $g(\tau(\tau_*))\equiv\sqrt{\frac{n(n-1)}{2}}\tanh^2\tau$ and defining $\psi\equiv \exp[\frac{1}{2}\int g(\tau_*)\dd{\tau_*}]f$, we find that \eqref{46} takes the Schr\"{o}dinger-like form
\begin{align}
-\dv[2]{\psi(\tau_*)}{\tau_*}+\left[\frac{1}{2}\dv{g(\tau_*)}{\tau_*}+\frac{1}{4}g^2(\tau_*)\right]\psi=\omega^2\psi(\tau_*), \label{47}
\end{align}
where the potential is $V=\frac{1}{2}\dv{g}{\tau_*}+\frac{1}{4}g^2$. Since $\tau_*=\sqrt{\frac{8}{{n(n-1)}}}\ln\sinh\tau$, we obtain
\begin{align}
g(\tau_*)=\sqrt{\frac{n(n-1)}{2}}\left(\exp[-{\sqrt{\frac{n(n-1)}{2}}\tau_*}]+1\right)^{-1},
\end{align}
and $V>0$ holds for all $\tau_*\in(-\infty,\infty)$, where $\tau_*=-\infty$ corresponds to the horizon ($\tau=0$) and $\tau_*=\infty$ to the spatial infinity ($\tau=\infty$).

To impose physical boundary conditions on \eqref{47}, note that $V(\tau_*=-\infty)=0$ and $V(\tau_*=\infty)=\frac{n(n-1)}{8}\equiv V_\infty$. At the horizon ($\tau_*=-\infty$), we require purely ingoing waves $\psi(\tau_*=-\infty)\sim e^{-i\omega \tau_*}$, and at infinity, either purely outgoing or damped solutions:
\begin{align}
\psi(\tau_*=\infty)\sim
\begin{cases}
e^{i\sqrt{\omega^2-V_\infty}\tau_*} &(\omega^2>V_\infty) \\
0 &(\mathrm{otherwise})
\end{cases}.
\end{align}
The first case satisfies $\Im(\omega)=0$ and hence is stable. Multiplying \eqref{47} by $\psi^*$ and integrating over $\tau_*\in(-\infty,\infty)$ with integration by parts yields
\begin{align}
\int_{-\infty}^\infty\left|\dv{\psi}{\tau_*}\right|^2\dd{\tau_*}+\int_{-\infty}^\infty V|\psi|^2\dd{\tau_*}=\omega^2\int_{-\infty}^\infty|\psi|^2\dd{\tau_*}+\left[\psi^*\dv{\psi}{\tau_*}\right]_{-\infty}^\infty. \label{48}
\end{align}
For $\psi(\tau_*=\infty)\sim 0$, the boundary term gives $\left[\psi^*\dv{\psi}{\tau_*}\right]_{-\infty}^\infty=i\omega$. Writing $\omega=\omega_\mathrm{R}-i\omega_\mathrm{I}$ $(\omega_\mathrm{R}>0)$ and noting that the left-hand side is real, we obtain
\begin{align}
-2i\omega_\mathrm{R}\omega_\mathrm{I}\int_{-\infty}^\infty|\psi|^2\dd{\tau_*}+i\omega_\mathrm{R}=0,\quad
\omega_\mathrm{I}=\frac{1}{2\int_{-\infty}^\infty|\psi|^2\dd{\tau_*}}\ge0\quad (\omega_\mathrm{R}\neq0),
\end{align}
indicating stability for $\omega_\mathrm{R}\neq0$. The case $\omega_\mathrm{R}=0$ implies $\omega^2\le0$, which contradicts $V>0$, and thus is excluded. Therefore, under the physical (quasinormal mode) boundary conditions, the positivity of $V$ ensures that no unstable mode with $\Im(\omega)<0$ appears, and the perturbation of the gauge field is stable.

\paragraph{\underline{The case $c_2=1$}}

This corresponds to the exact $\mathrm{AdS}_2$ limit of the solution in section \ref{subsubsec2.2.2}. Defining the tortoise coordinate $\tau_*$ via $\dv{\tau_*}{\tau}=\sqrt{\frac{8}{n(n-1)}}e^{-\Sigma}=\sqrt{\frac{8}{n(n-1)}}\frac{1}{\sinh2\tau}$ and noting that $\varphi'=-\frac{1}{2}\Sigma'$, we find that \eqref{27} becomes
\begin{align}
-\dv[2]{f}{\tau_*}+\left[-\frac{q^2}{8}\sinh^4\tau\right]f=\omega^2 f,
\end{align}
which takes the Schr\"{o}dinger form with potential $V=-\frac{q^2}{8}\sinh^4\tau$. The eigenvalue pair $(f,\omega)$ thus corresponds to bound states of $V$. Since $\tau_*=\sqrt{\frac{2}{n(n-1)}}\ln\tanh\tau$, the range $\tau\in[0,\infty)$ corresponds to $\tau_*\in[-\infty,0]$. Expressing $V$ in terms of $\tau_*$, we have
\begin{align}
V&=-\frac{q^2}{8}\sinh^4\tau \nonumber\\
&=-\frac{q^2}{8}\frac{1}{(e^{-\sqrt{2n(n-1)}\tau_*}-1)^2},
\end{align}
which monotonically decreases from $V=0$ at the horizon ($\tau_*=-\infty$) to $V=-\infty$ at infinity ($\tau_*=0$). Although it is difficult to solve the eigenvalue problem for $V$ exactly, the shape of $V$ suggests that there should be a bound state near the asymptotic region $\tau_* = 0$. Around $\tau_* \simeq 0$, we have $V \simeq -\frac{q^2}{16n(n-1)} \frac{1}{\tau_*^2}$, so that $V$ can be approximated by an inverse-square potential near infinity. In general, the eigenvalue problem for the inverse-square potential,
\begin{align}
-\dv[2]{r}\psi - \frac{\alpha}{r^2}\psi = E\psi
\end{align}
is known as the falling-to-center problem. Since the equation is invariant under the scale transformation $r \to \lambda r, \,\,\, E \to E/\lambda^2$, if a bound state exists, it follows that infinitely many bound states should exist as well, leading to a continuous energy spectrum unbounded from below. The condition for the existence of a bound state is $\alpha > \frac{1}{4}$, so in our system, an eigenvalue with $\omega^2 < 0$ appears when
\begin{align}
q^2 > 4n(n-1).
\end{align}
Thus, when the coupling constant is sufficiently large, an unstable mode localizes near infinity\footnote{Here, we have not explicitly specified boundary conditions, but since we focus on modes with $\Re(\omega)=0$, they are neither ingoing nor outgoing at the horizon and remain regular at infinity, satisfying the physical boundary conditions.} and grows as $w \propto e^{\abs{\omega}\tilde{t}}$. 

Consequently, since the black brane becomes unstable in the limit of decreasing temperature $T$, we conclude that a phase transition occurs at some point as the tempereture decreases, and the stable phase changes to another black brane configuration with a different gauge field profile.

\section{Another exact solution : non-rotating BTZ black hole}
\label{sectionBTZ}
In this section, we describe a method for deriving further exact solutions for $n<8$ within the setup where nontrivial gauge fields are placed on $S^n$, and provide an explicit example in which this procedure can indeed be carried out. In section \ref{section2}, for solutions in which $S^n$ is compactified with a constant radius, we were able to directly embed known solutions of two-dimensional heterotic supergravity. Motivated by this, let us now take some of the directions $(w_1, w_2, \cdots, w_k)$ from the (originally) flat space $\mathbb{R}^{8-n}$ and allow them to have nontrivial geometry, thereby enlarging the set of nontrivial directions to $(t, r, w_1, w_2, \cdots, w_k)$. 
Within this setup, if the radius $R$ of $S^n$ again becomes constant, in close analogy with the situation in section \ref{section2}, one obtains an effective $(k+2)$-dimensional heterotic supergravity action in which the combination $\mathcal{R} - \frac{\alpha^\prime}{2}\tr|F|^2$ plays the role of a negative cosmological constant. In other words, if an exact solution exists in $(k+2)$-dimensional heterotic supergravity, it can be directly embedded as a solution in the present higher-dimensional setup.

As an example, let us demonstrate that an exact solution can indeed be constructed in the case $k=1$. Among the ten-dimensional spacetime, we decompose the spatial $(n+1)$ dimensions into the radial direction $r$ and the $n$-dimensional sphere $S^n$. Of the remaining $(9-n)$ dimensions, we take one to be the time coordinate $t$ and one spatial coordinate to be $w$. The remaining $(7-n)$ dimensions are kept flat. Concretely, we adopt the ansatz\footnote{Here, we use the Lorentzian signature.}
\begin{align}
\dd{s}^2&=-e^{2\Sigma_1(r)}\dd{t}^2+e^{2\Sigma_2(r)}\dd{w^2}+\dd{r}^2+R^2(r)\dd{\Omega_n}^2+\dd{X^i}\dd{X^i}.
\end{align}
We place a gauge field on $S^n$ such that $\tr\abs{F} ^2=\frac{\mathsf{C}}{R^4}$, and in the present case we set the time component of the gauge field $A_t$ to zero.
Instead, we introduce a $B$-field whose nonvanishing components are restricted to the $(t,w,r)$ directions, namely
$B_{ab}$ with $a,b=t,r,w$.
As a remark, although we choose a special gauge field configuration on $S^n$ so that the $B$-field on $S^n$ vanishes, this condition applies only on $S^n$; the $B$-field components along other directions are allowed to be nonzero. The action in this case given by
\begin{align}
S=\int\dd{x^{10}}\sqrt{-G}\, e^{-2\Phi}\left(\mathcal{R}+4(\nabla\Phi)^2-\frac{\alpha^\prime}{2}\tr\abs{F}^2-\frac{1}{12}H_{abc}H^{abc}\right),
\end{align}
where $H=\dd{B}$. Performing the ADM decomposition with respect to the $t$ and $w$ directions in the $(n+3)$-dimensional spacetime excluding the flat part, we find that the curvature $\mathcal{R}$ can be written as
\begin{align}
\mathcal{R}=\mathcal{R}_{n+1}+(\text{terms independent of } R).
\label{Rdecomp}
\end{align}
Since $\mathcal{R}_{n+1}$ can be written as\footnote{In detail, see \eqref{FLRW} in Appendix \ref{AppendixA}.}
\begin{align}
\mathcal{R}_{n+1}=-2n\dv{r}\left(\dv{\ln R}{r}\right)-n(n+1)\left(\dv{\ln R}{r}\right)^2+\frac{n(n-1)}{R^2}, 
\end{align}
the terms involving $R$ appearing in the action are the same as those in section \ref{section2}. Therefore, by setting $R=\ell_0 e^{\sigma(r)}$, the equation obtained from varying with respect to $\sigma$ is given by \eqref{13}, and $\sigma=0$ is indeed a solution.
In this situation, we find $R=\mathrm{const}$, and as in section \ref{section2} the combination
$\mathcal{R}_{S^n}-\frac{\alpha^\prime}{2}\tr\abs{F}^2$
plays the role of a negative cosmological constant, reducing the theory to a three-dimensional spacetime theory in the $(t,r,w)$ directions. Namely, with
$\Lambda=-\left(\mathcal{R}_{S^n}-\frac{\alpha^\prime}{2}\tr\abs{F}^2 \right)$,
the effective action can be written as
\begin{align}
S=\int\dd^3{x}\sqrt{-G_3}\, e^{-2\Phi}\left(\mathcal{R}_3-\Lambda+4(\nabla\Phi)^2-\frac{1}{12}H_{abc}H^{abc}\right). \label{3Daction}
\end{align}
This system was analyzed in \cite{Horowitz:1993jc}, where it was shown that the BTZ black hole (\cite{Banados:1992wn}) is a solution. Since we have assumed a spherically symmetric metric, it follows  that the spherically symmetric BTZ black hole, namely the $\mathrm{AdS}_3$--Schwarzschild black hole, is realized as a solution. Explicitly, defining $l^2=-4/\Lambda$, we find the solution given by 
\begin{align}
\dd{s}^{2}
= -\left(\frac{\hat r^{2}}{l^{2}}-1\right)\dd\hat{t}^{2}
+&\left(\frac{\hat r^{2}}{l^{2}}-1\right)^{-1}\dd\hat r^{2}
+\hat r^{2}\dd\hat\varphi^{2}+\ell_0^2\dd{\Omega_n}^2+\dd{X^i}\dd{X^i},  \\
&B_{\hat{\varphi}\hat{t}}=\frac{\hat{r}^2}{l},  \\
&\Phi=0.
\end{align}
Here the hatted coordinates denote a coordinate system different from the previously used coordinates such as $t$ and $r$. 

This is a non-rotating BTZ black hole. 
Whether the rotating BTZ black hole is also realized as a solution in ten dimensions requires a more detailed analysis. Indeed, in the absence of spherical symmetry, the curvature $\mathcal{R}$ can no longer be decomposed as in \eqref{Rdecomp}, and it is therefore nontrivial whether a solution with constant $S^n$ radius (i.e. $\sigma = 0$) still exists.

\section{Summary and discussion}\label{section4}
In this section, we summarize the results obtained in this paper, prompt the readers to a possible phase diagram of black branes, and discuss several related issues.

In this work, we constructed a new exact black brane solution in heterotic supergravity. This solution generalizes the known charged black brane solution \cite{3} which is obtained by decomposing the $(n+1)$ spatial dimensions of the 10-dimensional spacetime into a radial direction $r$ and an $n$-sphere $S^n$ and placing a nontrivial gauge field on $S^n$. In section \ref{section2}, by further introducing a time component $A_t$ of the gauge field, we obtained our new solution \eqref{exact_solution}. This charged blackbrane solution is exact and parameterized by a single real parameter $c_2$ ($0\leq c_2 \leq 1$). The physical quantities of our black brane, such as change, mass and temperature, are given by this parameter $c_2$.

Interestingly, this solution reduces to an exact $\mathrm{AdS}_2$ geometry when the parameter $c_2$ is set to 1. This is peculiar in that, unlike the case of the near-horizon limit of an extremal Reissner--Nordstr\"{o}m black hole where $\mathrm{AdS}_2$ emerges only after taking a near-horizon limit, the $\mathrm{AdS}_2$ structure here appears without any limiting the region\footnote{On the other hand, the system considered here admits not only the solution with $\sigma(\tau)=0$ but also solutions with $\sigma(\tau)\neq 0$. In \cite{3}, it was numerically suggested (for the $A_{t_\mathrm{E}}=0$ system) that the $\sigma(\tau)=0$ solution effectively corresponds to the near-horizon limit of solutions with $\sigma(\tau)\neq 0$. In this sense, the present $\mathrm{AdS}_2$ geometry may be interpreted as the near-horizon limit of solutions with $\sigma(\tau)\neq 0$.}. Moreover, we have shown that the two-dimensional $(t,r)$ sector of this new black brane shares the same structure as the solution derived in noncritical two-dimensional heterotic string theory. This is because the curvature of the compactified $S^n$ and the gauge field placed on $S^n$ play the role of the central charge in the two-dimensional heterotic string theory. This observation points to a new possibility of embedding the spacetime that appears in noncritical superstring theory into ten-dimensional superstring theory. From this perspective, in section \ref{sectionBTZ} we propose a method for deriving new exact solutions in ten-dimensional heterotic supergravity, and successfully construct explicit examples based on this approach. The exact solution obtained in this way contains a three-dimensional sector described by a non-rotating BTZ black hole and is supported by a nontrivial $B$-field. 

Furthermore, in section \ref{section3}, we analyzed gauge field perturbations of this new black brane solution for $n=2, 4$ and showed that unstable modes appear as the temperature is lowered. The peak of the unstable mode is localized in the asymptotic region, and as the charge increases, the gauge field gradually condenses from far away, eventually indicating a transition to some unknown black brane solution with the gauge field condensation.

The perturbation analysis is largely benefitted from the fact that we obtained an exact solution. Therefore, 
our results clearly demonstrate the effectiveness of the research strategy of explicitly constructing exact black hole solutions with multiple gauge fields and probing their dynamics and phase structure through perturbative analysis. This work provides a concrete example toward understanding the stability and phase transitions of black holes arising in heterotic string theory.

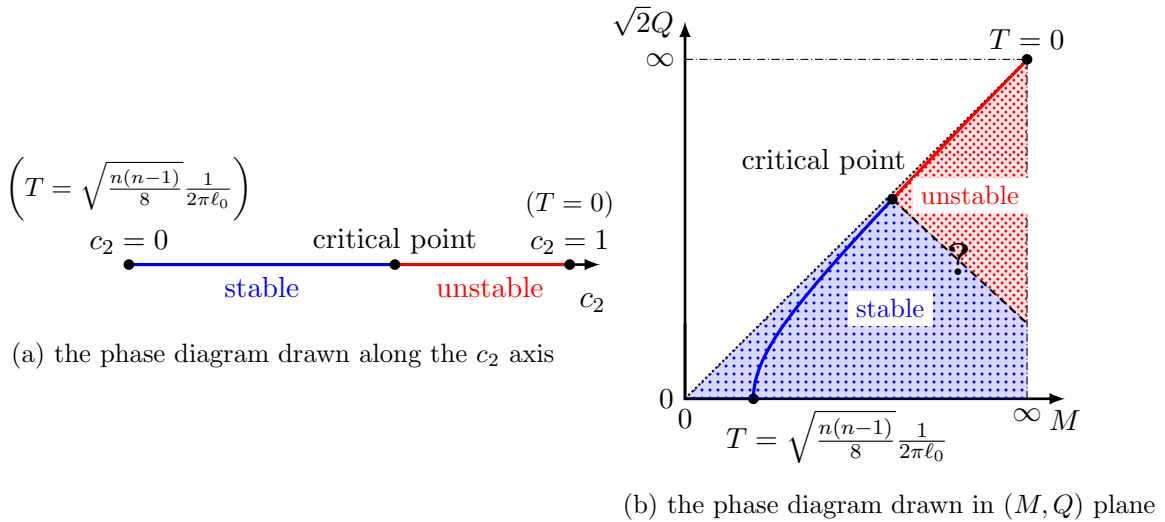
\begin{figure}
\begin{tabular}{cc}
\begin{minipage}[c]{0.5\columnwidth}
\begin{tikzpicture}[line width=1pt,>=latex]
\draw[->] (-3,0)--(3.2,0);
\node at (3.1,-0.5) {\large{$c_2$}};
\draw[very thick,blue] (-3,0)--(0.5,0);
\draw[very thick,red] (0.5,0)--(2.8,0);
\fill (-3,0)node[above]{$c_2=0$} circle [radius=2pt];
\fill (0.5,0)node[above]{critical point} circle [radius=2pt];
\fill (2.8,0)node[above]{$c_2=1$} circle [radius=2pt];
\node at (-1.25,-0.3) {{\color{blue}stable}};
\node at (1.75,-0.3) {{\color{red}unstable}};
\node at (-3,1) {\small{$\left(T=\sqrt{\frac{n(n-1)}{8}}\frac{1}{2\pi\ell_0}\right)$}};
\node at (2.8,0.8) {\small{$(T=0)$}};
\end{tikzpicture}
\subcaption{the phase diagram drawn along the $c_2$ axis}
\label{fig3}
\end{minipage} &

\begin{minipage}[c]{0.5\columnwidth}
\begin{tikzpicture}[>=latex]
\draw[->,line width=1pt] (0,1)--(0,0)node[below]{$0$}--(5,0) node[below] {$M$};
\draw[->,line width=1pt] (1,0)--(0,0)node[left]{$0$}--(0,5) node[left] {$\sqrt{2}Q$};
\draw[densely dotted,line width=0.8pt] (0,0)--(4.5,4.5);
\draw[densely dash dot] (0,4.5) node[left]{$\infty$}--(4.5,4.5);
\draw[densely dash dot] (4.5,0) node[below]{$\infty$}--(4.5,4.5);
\draw[domain=0:2.2, smooth, samples=200, thick]
  plot ({cosh(\x)-0.1}, {sinh(\x)});
\draw[domain=0:1.7, smooth, samples=200, very thick, blue]
  plot ({cosh(\x)-0.1}, {sinh(\x)});
\draw[domain=1.7:2.2, smooth, samples=200, very thick, red]
  plot ({cosh(\x)-0.1}, {sinh(\x)});
\node at ({cosh(1.7)-1},{sinh(1.7)+0.5}) {critical point};
\fill (0.9,0) circle [radius=2pt];
\node at (2,-0.5){$T=\sqrt{\frac{n(n-1)}{8}}\frac{1}{2\pi\ell_0}$};
\fill (4.5,4.5)node[above]{$T=0$} circle [radius=2pt];
\fill[pattern=dots, pattern color=blue] (0,0)--(4.5,0)--(4.5,1)--(2.685,2.685)--(0,0);
\fill[pattern=crosshatch dots, pattern color=red] (4.5,1)--(4.5,4.5)--(2.685,2.685)--(4.5,1);
\filldraw[fill=blue, opacity=0.15](0,0)--(4.5,0)--(4.5,1)--(2.685,2.685)--(0,0);
\filldraw[fill=red, opacity=0.1](4.5,1)--(4.5,4.5)--(2.685,2.685)--(4.5,1);
\node at (2.7,1.2) {\colorbox{white}{{\small{\color{blue}stable}}}};
\node at (3.75,2.7) {\colorbox{white}{\small{\color{red}unstable}}};
\draw[densely dashed,thick] ({cosh(1.7)-0.1},{sinh(1.7)})--(4.5,1);
\fill ({cosh(1.7)-0.1},{sinh(1.7)}) circle [radius=2pt];
\node at (3.6,1.85) {\LARGE{\textbf{?}}};

\end{tikzpicture}
\subcaption{the phase diagram drawn in $(M,Q)$ plane}
\label{fig4}
\end{minipage}

\end{tabular}
\caption{the rough phase diagrams of the new black branes}
\end{figure}

Here, based on the results obtained in section \ref{section3}, let us draw a phase diagram of this black brane. Of course, since we have examined the stability only for the no-charge solution ($c_2=0$) and for the extremal solution ($c_2=1$), we do not know in detail where the phase transition occurs. However, because we know that a phase transition must occur somewhere in between, we can still draw a qualitative diagram. Since the temperature $T$ is determined solely by the parameter $c_2$, the parameter region in $c_2$ where the solution is stable can be depicted as in Figure \ref{fig3}. Let us further extend the phase diagram to that in terms of the black brane charge $Q$ and mass $M$, a more popular form of phase diagrams. The quantities $Q$ and $M$ are given by \eqref{charge} and \eqref{mass}, and the possible values of $(M,Q)$ trace a certain curve on the $(M,Q)$ plane when $c_2$ is treated as a parameter. By rewriting the parameter as $c_2 = \tanh \gamma$ with $\gamma \in [0,\infty]$, we obtain
\begin{align}
M&= {A_nV_p}\sqrt{\frac{n(n-1)}{2\ell_0^2}}\, e^{-2\Phi_0} \cosh 2\gamma ,\\
\sqrt{2}\,Q&= {A_nV_p}\sqrt{\frac{n(n-1)}{2\ell_0^2}}\, e^{-2\Phi_0} \sinh 2\gamma ,
\end{align}
from which it follows that $(M,Q)$ traces a hyperbola for fixed $\Phi_0$. It should be emphasized that the fact that the mass $M$ and charge $Q$ of the black brane are not independent but determined by a single parameter is a consequence of our focus on the solution with $\sigma(\tau)=0$ to look for exact solutions. If one instead study numerically more general solutions with nontrivial $\sigma(\tau)$, one would be able to obtain a family of solutions in which $M$ and $Q$ become independent quantities. Those solutions would occupy regions of the $(M,Q)$ plane including our hyperbola, and it is expected that the discussion of instability can be extended to the broad regions as well. Therefore, the phase diagram with fixed $\Phi_0$ on the $(M,Q)$ plane would look qualitatively like Figure \ref{fig4}. We have not plotted anything in the region $\sqrt{2}Q>M$, which can be regarded as superextremal on the phase diagram. By being able to draw the phase diagram for the solution with $\sigma(\tau)=0$, we can roughly predict the stability of solutions with $\sigma(\tau)\neq 0$, which would otherwise be accessible only through numerical analysis. This provides a valuable clue toward understanding the dynamics of the system.

Below, we discuss both the strengths and subtle points of the newly obtained black brane solution.

\begin{itemize}
\item \textbf{Possibility of other solutions}

Using the method employed in this paper--namely, placing a nontrivial gauge field on $S^n$ to compactify the sphere--one can embed previously studied solutions of heterotic supergravity directly into our setup. For example, in section \ref{section2} we showed that exact solutions of two-dimensional heterotic supergravity can be nontrivially embedded into ten-dimensional heterotic supergravity, while in section \ref{sectionBTZ} we demonstrated that the same is true for solutions of three-dimensional heterotic supergravity. By applying this strategy, one can expect to construct many more exact solutions in ten dimensions. More concretely, one may take $k$ directions from the flat $\mathbb{R}^{8-n}$ sector and reduce the system to $(k+2)$-dimensional heterotic supergravity. In this procedure, as discussed in section \ref{sectionBTZ}, the method is guaranteed to work for spherically symmetric solutions, since a solution with $R=\mathrm{const}$ always exists. By contrast, in situations with rotation, one must explicitly verify whether a solution with $R=\mathrm{const}$ is present. For exact solutions of heterotic supergravity in noncritical dimensions, see \cite{Horowitz:1992jp, Horne:1991gn, Horne:1991cn} for a detailed discussion.

We emphasize the nontriviality of our new black brane solution in comparison to the solutions obtained in two-dimensional heterotic supergravity (\cite{McGuigan:1991qp, Gibbons:1992rh}). The charged black holes studied in two dimensions lack an antisymmetric $B$-field term because of the dimensionality, and their cosmological constant arises from the central charge of the anomalous dimension. By contrast, the new solution considered in this paper eliminates the $B$-field on $S^n$ by choosing a special gauge field configuration (for example, an equal number of instantons and anti-instantons for $n=4$), and the cosmological constant originates from the curvature of $S^n$ and the gauge fields on it. Therefore, the mechanisms of getting the solutions are entirely different, although the 2-dimensional parts of the solutions are shared. In fact, our result of the stability analysis does not apply to the two-dimensional solutions since the perturbation modes do not live in the two dimensions, and it has been shown that no unstable perturbative modes arise in two dimensions
\cite{Ahn:1992df, Li:2001ct, Xi:2005stw}.

\item \textbf{Whether supersymmetry is preserved}

Our new black brane solution has the property that its $(t,r)$ sector coincides with the exact solutions previously found in two-dimensional heterotic supergravity (\cite{McGuigan:1991qp}, \cite{Gibbons:1992rh}). In the two-dimensional setup, when $c_2 = 1$, namely in the zero-temperature limit $T = 0$, the spacetime becomes exactly $\mathrm{AdS}_2$. Although we have not yet verified it through an explicit computation, the solution at $c_2 = 1$ appears to preserve supersymmetry. Indeed, in this limit the dilaton becomes constant and the spacetime isometry is enhanced to $SO(1,2)$, which suggests that the enhancement originates from supersymmetry.

However, when this solution is viewed as a ten-dimensional spacetime, the situation is different. The perturbative analysis of the gauge field carried out in section \ref{section3} shows that an instability appears at $c_2 = 1$, which indicates that supersymmetry is not preserved in ten dimensions. In other words, while the solution seems supersymmetric from the two-dimensional viewpoint, it loses supersymmetry when embedded in ten dimensions.

This discrepancy can be understood as follows. In ten dimensions, the directions other than $(t,r)$ form $S^n \times \mathbb{R}^p$, and a nontrivial gauge field configuration exists on $S^n$. Most likely, the supersymmetry transformations involving the metric components and the gauge field on $S^n$ do not leave the solution invariant, resulting in supersymmetry breaking in ten dimensions. To confirm this rigorously, one would need to explicitly write down the gauge field on $S^n$ and evaluate the supersymmetry transformations for all fields. This task is beyond the scope of the present work and is left for future investigation.

\item \textbf{Toward heterotic compactification to the Standard Model}

Among the black brane solutions obtained in this work and the underlying FKWY black brane solutions, the case $n = 4$ is of particular interest as a possible framework for deriving the Standard Model from heterotic string theory through a new mechanism of the compactification. In the case $n=4$, the noncompact part includes $\mathbb{R}^4$, and by Wick-rotating one of its directions, one can obtain Minkowski spacetime $\mathbb{R}^{1,3}$. We also Wick-rotate the time direction of the original black brane to its Euclidean time. Since the $S^4$ part of the solution is compactified at roughly the string scale, only the $r$ direction and $\mathbb{R}^{1,3}$ are effectively observable at low energies. Therefore, if gauge fields or matter fields localize in some region along the $r$ direction, it may be possible to construct the Standard Model on $\mathbb{R}^{1,3}$. If such a picture were realized, it would offer an attractive analytic model, free from the computational complexity of Calabi--Yau compactification and describable entirely in terms of elementary functions.

However, the actual situation is not so simple. Taking the FKWY black brane as a background and considering perturbations of $E_6$ or $SO(10)$ gauge fields living on it, one finds that no normalizable modes exist. Thus, gauge fields do not localize at any particular point along the $r$ direction, and the expected brane-world scenario cannot be realized. The origin of this non-normalizability lies in the behavior of the dilaton $\Phi$, which diverges to $-\infty$ in the asymptotic region as shown in \eqref{10}. As a result, when the perturbative solution is substituted into the action $S$, the divergence in $e^{-2\Phi}$ remains un-canceled.

In the new black brane solution constructed in this paper, the dilaton becomes constant when $c_2 = 1$, which might appear to alleviate this issue. However, If one attempts to directly use the solutions obtained in this paper for the Standard Model, serious problems arise. 
In particular, the gauge field $A_{t_\mathrm{E}}$ becomes imaginary, leading to non-unitarity. 
One might think that this problem could be avoided by choosing the integration constant $iC$ in \eqref{19} to be real, i.e.\ by taking $C$ to be purely imaginary, so that $A_{t_\mathrm{E}}$ becomes real. 
However, this choice implies $C^2<0$, which completely changes the structure of the subsequent equations of motion. 
As a consequence, the solutions are qualitatively different, and in particular no solution with a constant dilaton can be obtained in this case. 

From these considerations, it is clear that the new black brane solutions obtained in this paper cannot be directly used for string compactification. 
Therefore, one must either find a mechanism to control the divergence of $e^{-2\Phi}$ in the FKWY black brane solutions, or discover alternative black brane solutions with more suitable physical properties for string compactification.

\end{itemize}

The endeavor to construct new black hole solutions within heterotic string theory is of fundamental importance, both for deepening our understanding of the spacetime dynamics emerging from string theory and for developing new frameworks that may ultimately lead to the realization of the Standard Model. Even within the class of systems investigated in this work, many spatial directions have been left flat, indicating that there remains considerable room for discovering further exact solutions.

While it is widely believed that quantum aspects of gravity become prominent only at the Planck scale, it is equally plausible that the imprint of quantum gravity may already be encoded in classical solutions of low-energy effective theories. One of the aims of exploring classical solutions in the framework of supergravity is to uncover such configurations that carry the signatures of quantum gravity.
Our exploration revealed a nobel phase diagram of black brane solutions, which expands the string theory landscape. The endeavor thereby pushes us closer to the true nature of quantum gravity.


\acknowledgments

We are grateful to Y.~Abe and K.~Yoshida for valuable comments. 
S.~C.~would like to thank R.~Adachi and Y.~Gen for helpful discussions. 
The work of K.~H.~was supported in part by JSPS KAKENHI Grant No.~JP22H01217, JP22H05111 and JP22H05115.


\appendix

\section{Derivation of action and equations of motion} \label{AppendixA}
In this appendix, we derive the explicit form of the action $S$ and the corresponding equations of motion (EOM) that were omitted in Section \ref{section2} and \ref{section3}. In the following derivation, we consider the case including the gauge field $A_{\tilde{t}_\mathrm{E}}$, while the results in Section \ref{section2} can be obtained simply by setting $A_{\tilde{t}_\mathrm{E}}=0$.

Assuming the metric
\begin{align}
\dd{s}^2&=e^{2\Sigma(\tau)}\dd{t_{\mathrm{E}}^2}+N^2(r)\dd{r}^2+R^2(r)\dd{\Omega_n^2}+\dd{X^i}\dd{X^i} \\
&=\ell_0^2e^{2\Sigma(\tau)}\dd{\tilde{t}_\mathrm{E}^2}+\frac{8\ell_0^2}{n(n-1)}N^2(\tau)\dd{\tau^2}+\ell_0^2e^{2\sigma(\tau)}\dd{\Omega_n^2}+\dd{X^i}\dd{X^i},
\end{align}
the action becomes
\begin{align}
S&=V_p\int\dd^{n+2}{x}\sqrt{G}e^{-2\Phi}\left(\mathcal{R}+4(\nabla\Phi)^2-\frac{\alpha^\prime}{2}\left[\frac{\mathsf{C}}{R^4}+\frac{1}{q^2}\frac{2}{\alpha^\prime}\frac{n(n-1)}{\ell_0^2}A_{\tilde{t}_{\mathrm{E}}}^{\prime 2}e^{-2\Sigma}\right]\right) \\
&\propto \int\dd^{n+1}{x}\sqrt{G_{n+1}}e^{-2\Phi+\Sigma} \notag\\
&\qquad \left(\mathcal{R}_{n+1}-2e^{-\Sigma}\nabla^2_{n+1}e^{\Sigma}+4(\nabla\Phi)^2-\frac{n(n-1)}{2\ell_0^2}e^{-4\sigma}-\frac{1}{q^2}\frac{n(n-1)}{\ell_0^2}A_{\tilde{t}_{\mathrm{E}}}^{\prime 2}e^{-2\Sigma}\right).
\end{align}
Here, $\mathcal{R}_{n+1}$ is the scalar curvature of the $(n+1)$-dimensional spatial part after decomposing the $(n+2)$-dimensional spacetime as
\begin{align}
\mathcal{R}=\mathcal{R}_{n+1}-2e^{-\Sigma}\nabla^2_{n+1}e^{\Sigma},
\end{align}
where $G_{n+1,\mu\nu}$ denotes the $(n+1)$-dimensional metric, $\nabla$ is the covariant derivative with respect to $G_{\mu\nu}$, and $\nabla_{n+1}$ is that with respect to $G_{n+1,\mu\nu}$.  
Defining $\Phi_{n+1}=\Phi-\frac{1}{2}\Sigma$, we have
\begin{align}
&\int\dd^{n+1}{x}\sqrt{G_{n+1}}e^{-2\Phi_{n+1}}\left(-2e^{-\Sigma}\nabla^2_{n+1}e^{\Sigma}+4(\nabla\Phi)^2\right) \nonumber\\
=&\int\dd^{n+1}{x}\sqrt{G_{n+1}}e^{-2\Phi_{n+1}}\left(-2\nabla_{n+1}^2\Sigma-2(\nabla_{n+1}\Sigma)^2+4\left(\nabla_{n+1}\Phi_{n+1}+\frac{1}{2}\nabla_{n+1}\Sigma\right)^2\right)\nonumber\\
=&\int\dd^{n+1}{x}\sqrt{G_{n+1}}e^{-2\Phi_{n+1}}\left(4(\nabla_{n+1}\Phi)^2-(\nabla_{n+1}\Sigma)^2+4(\nabla_{n+1}^\mu\Phi_{n+1})(\nabla_{n+1,\mu}\Sigma)-2\nabla_{n+1}^2\Sigma\right)\nonumber\\
=&\int\dd^{n+1}{x}\sqrt{G_{n+1}}e^{-2\Phi_{n+1}}\left(4(\nabla_{n+1}\Phi)^2-(\nabla_{n+1}\Sigma)^2\right) \notag\\
&\qquad -2\int\dd^{n+1}{x}\sqrt{G_{n+1}}\nabla_\mu\left(e^{-2\Phi_{n+1}}\nabla_{n+1}^\mu\Sigma\right).
\end{align}
Neglecting the total derivative term, the action becomes
\begin{align}
S\propto\int&\dd^{n+1}{x}\sqrt{G_{n+1}}e^{-2\Phi_{n+1}} \notag\\
&\left(\mathcal{R}_{n+1}+4(\nabla_{n+1}\Phi_{n+1})^2-(\nabla_{n+1}\Sigma)^2-\frac{n(n-1)}{2\ell_0^2}e^{-4\sigma}-\frac{1}{q^2}\frac{n(n-1)}{\ell_0^2}A_{\tilde{t}_{\mathrm{E}}}^{\prime 2}e^{-2\Sigma}\right).
\end{align}
Assuming spherical symmetry, we can use the standard FLRW-type expression for $\mathcal{R}_{n+1}$ (interpreting the time coordinate $t$ as the radial coordinate $r$):
\begin{align}
\mathcal{R}_{n+1}=-2n\frac{1}{N}\dv{r}\left(\frac{1}{N}\dv{\ln R}{r}\right)-n(n+1)\left(\frac{1}{N}\dv{\ln R}{r}\right)^2+\frac{n(n-1)}{R^2}. \label{FLRW}
\end{align}
Substituting this into the above expression, setting $R=\ell_0e^\sigma$, $r=\sqrt{\frac{8}{n(n-1)}}\ell_0\tau$, and $\varphi=\Phi_{n+1}-\frac{n}{2}\sigma$, and taking the gauge $N=1$, we obtain
\begin{align}
S&\propto \int\dd{\tau}e^{-2\varphi}\left(-2n\sigma^{\prime\prime}+4n\sigma^\prime\varphi^\prime-n\sigma^{\prime 2}+4\varphi^{\prime 2}-\Sigma^{\prime 2}+8e^{-2\sigma}-4e^{-4\sigma}-\frac{8}{q^2}A^{\prime 2}_{\tilde{t}_{\mathrm{E}}}e^{-2\Sigma}\right) \nonumber\\
&=\int\dd{\tau}e^{-2\varphi}\left(-n\sigma^{\prime 2}+4\varphi^{\prime 2}-\Sigma^{\prime 2}+8e^{-2\sigma}-4e^{-4\sigma}-\frac{8}{q^2}A^{\prime 2}_{\tilde{t}_{\mathrm{E}}}e^{-2\Sigma}\right) \notag \\
&\qquad +\int\dd{\tau}\left[-e^{-2\varphi}(2n\sigma^\prime)\right]^{\prime}.
\end{align}
Neglecting the total derivative term, we obtain the action
\begin{align}
S&\propto\int\dd{\tau}e^{-2\varphi}\left(-\frac{n}{8}\sigma^{\prime 2}-\frac{1}{8}\Sigma^{\prime 2}+\frac{1}{2}\varphi^{\prime 2}+e^{-2\sigma}-\frac{1}{2}e^{-4\sigma}-\frac{1}{q^2}A^{\prime 2}_{\tilde{t}_{\mathrm{E}}}e^{-2\Sigma}\right) \\
&\equiv\int\dd{\tau}\mathcal{L}.
\end{align}
In the canonical formalism with respect to $\tau$, the conjugate momenta for each variable are
\begin{align}
&\Pi_\sigma=\pdv{\mathcal{L}}{\sigma^\prime}=-\frac{n}{4}\sigma^\prime e^{-2\varphi}, \\
&\Pi_\varphi=\pdv{\mathcal{L}}{\varphi^\prime}=\varphi^\prime e^{-2\varphi}, \\
&\Pi_\Sigma=\pdv{\mathcal{L}}{\Sigma^\prime}=-\frac{1}{4}\Sigma^\prime e^{-2\varphi}, \\
&\Pi_{A_{\tilde{t}_{\mathrm{E}}}}=\pdv{\mathcal{L}}{A_{\tilde{t}_{\mathrm{E}}}}=-\frac{2}{q^2}A_{\tilde{t}_{\mathrm{E}}}^\prime e^{-2\varphi-2\Sigma}.
\end{align}
Then the Hamiltonian density $\mathcal{H}$ is
\begin{align}
\mathcal{H}&\equiv \Pi_{\sigma}\sigma^\prime+\Pi_\varphi\varphi^\prime+\Pi_\Sigma\Sigma^\prime+\Pi_{A_{\tilde{t}_{\mathrm{E}}}}A_{\tilde{t}_{\mathrm{E}}}^\prime-\mathcal{L} \nonumber\\
&=e^{2\varphi}\left(-\frac{2}{n}\Pi_\sigma^2-2\Pi_\Sigma^2+\frac{1}{2}\Pi_\varphi^2-\frac{q^2}{4}e^{2\Sigma}\Pi_{A_{\tilde{t}_{\mathrm{E}}}}^2\right)
-e^{-2\varphi}\left(e^{-2\sigma}-\frac{1}{2}e^{-4\sigma}\right).
\end{align}
From the canonical equations, we obtain the equations of motion:
\begin{align}
&\frac{n}{4}(\sigma^{\prime\prime}-2\varphi^\prime\sigma^\prime)-2(e^{-2\sigma}-e^{-4\sigma})=0, \label{sigmaeq}\\
&\Sigma^{\prime\prime}-2\varphi^\prime\Sigma^\prime=-\frac{8}{q^2}e^{-2\Sigma}A^{\prime 2}_{\tilde{t}_\mathrm{E}}, \\
&\varphi^{\prime\prime}-\varphi^{\prime2}-\frac{n}{4}\sigma^{\prime 2}-\frac{1}{4}\Sigma^{\prime 2}+2e^{-2\sigma}-e^{-4\sigma}=0, \\
&A^{\prime\prime}_{\tilde{t}_\mathrm{E}}-2(\varphi^\prime+\Sigma^\prime)A^{\prime}_{\tilde{t}_\mathrm{E}}=0.
\end{align}
In addition, the constraint condition in the canonical formalism, corresponding to $N=1$, is given by $\mathcal{H}=0$:
\begin{align}
\mathcal{H}&=e^{2\varphi}\left(-\frac{2}{n}\Pi_\sigma^2-2\Pi_\Sigma^2+\frac{1}{2}\Pi_\varphi^2-\frac{q^2}{4}e^{2\Sigma}\Pi_{A_{\tilde{t}_{\mathrm{E}}}}^2\right)
-e^{-2\varphi}\left(e^{-2\sigma}-\frac{1}{2}e^{-4\sigma}\right) \nonumber\\
&=\frac{n}{4}\sigma^{\prime 2}+\frac{1}{4}\Sigma^{\prime 2}-\varphi^{\prime 2}+2e^{-2\sigma}-e^{-4\sigma}=0.
\end{align}

\bibliographystyle{JHEP}
\bibliography{biblio.bib}

@article{Gregory:1993vy,
    author = "Gregory, R. and Laflamme, R.",
    title = "{Black strings and p-branes are unstable}",
    eprint = "hep-th/9301052",
    archivePrefix = "arXiv",
    doi = "10.1103/PhysRevLett.70.2837",
    journal = "Phys. Rev. Lett.",
    volume = "70",
    pages = "2837--2840",
    year = "1993"
}

@article{Gregory:1994bj,
    author = "Gregory, Ruth and Laflamme, Raymond",
    title = "{The Instability of charged black strings and p-branes}",
    eprint = "hep-th/9404071",
    archivePrefix = "arXiv",
    reportNumber = "DAMTP-R-94-7, LA-UR-93-4473",
    doi = "10.1016/0550-3213(94)90206-2",
    journal = "Nucl. Phys. B",
    volume = "428",
    pages = "399--434",
    year = "1994"
}

@article{Kol:2004ww,
    author = "Kol, Barak",
    title = "{The Phase transition between caged black holes and black strings: A Review}",
    eprint = "hep-th/0411240",
    archivePrefix = "arXiv",
    doi = "10.1016/j.physrep.2005.10.001",
    journal = "Phys. Rept.",
    volume = "422",
    pages = "119--165",
    year = "2006"
}

@article{Harmark:2007md,
    author = "Harmark, Troels and Niarchos, Vasilis and Obers, Niels A.",
    title = "{Instabilities of black strings and branes}",
    eprint = "hep-th/0701022",
    archivePrefix = "arXiv",
    reportNumber = "CPHT-RR114-1206",
    doi = "10.1088/0264-9381/24/8/R01",
    journal = "Class. Quant. Grav.",
    volume = "24",
    pages = "R1--R90",
    year = "2007"
}

@article{Gubser:2000ec,
    author = "Gubser, Steven S. and Mitra, Indrajit",
    title = "{Instability of charged black holes in Anti-de Sitter space}",
    eprint = "hep-th/0009126",
    archivePrefix = "arXiv",
    reportNumber = "PUPT-1953",
    journal = "Clay Math. Proc.",
    volume = "1",
    pages = "221",
    year = "2002"
}

@article{Gubser:2000mm,
    author = "Gubser, Steven S. and Mitra, Indrajit",
    title = "{The Evolution of unstable black holes in anti-de Sitter space}",
    eprint = "hep-th/0011127",
    archivePrefix = "arXiv",
    reportNumber = "PUPT-1966",
    doi = "10.1088/1126-6708/2001/08/018",
    journal = "JHEP",
    volume = "08",
    pages = "018",
    year = "2001"
}

@article{Gubser:2004dr,
    author = "Gubser, Steven S.",
    title = "{The Gregory-Laflamme instability for the D2-D0 bound state}",
    eprint = "hep-th/0411257",
    archivePrefix = "arXiv",
    reportNumber = "PUPT-2144",
    doi = "10.1088/1126-6708/2005/02/040",
    journal = "JHEP",
    volume = "02",
    pages = "040",
    year = "2005"
}

@article{Reall:2001ag,
    author = "Reall, Harvey S.",
    title = "{Classical and thermodynamic stability of black branes}",
    eprint = "hep-th/0104071",
    archivePrefix = "arXiv",
    reportNumber = "QMUL-PH-01-06",
    doi = "10.1103/PhysRevD.64.044005",
    journal = "Phys. Rev. D",
    volume = "64",
    pages = "044005",
    year = "2001"
}

@article{Straumann:1990as,
    author = "Straumann, N. and Zhou, Z. H.",
    title = "{Instability of a colored black hole solution}",
    doi = "10.1016/0370-2693(90)90951-2",
    journal = "Phys. Lett. B",
    volume = "243",
    pages = "33--35",
    year = "1990"
}

@article{Brodbeck:1994vu,
    author = "Brodbeck, Othmar and Straumann, Norbert",
    title = "{Instability proof for Einstein Yang-Mills solitons and black holes with arbitrary gauge groups}",
    eprint = "gr-qc/9411058",
    archivePrefix = "arXiv",
    reportNumber = "ZU-TH-38-94",
    doi = "10.1063/1.531441",
    journal = "J. Math. Phys.",
    volume = "37",
    pages = "1414--1433",
    year = "1996"
}

@article{Zhou:1992sb,
    author = "Zhou, Z. H.",
    title = "{Instability of SU(2) Einstein Yang-Mills solitons and nonAbelian black holes}",
    journal = "Helv. Phys. Acta",
    volume = "65",
    pages = "767--819",
    year = "1992"
}

@article{Straumann:1989tf,
    author = "Straumann, Norbert and Zhou, Zhi-Hong",
    title = "{Instability of the Bartnik-mckinnon Solution of the Einstein {Yang-Mills} Equations}",
    reportNumber = "PRINT-89-0962 (ZURICH)",
    doi = "10.1016/0370-2693(90)91188-H",
    journal = "Phys. Lett. B",
    volume = "237",
    pages = "353--356",
    year = "1990"
}

@article{Polchinski:2003bq,
    author = "Polchinski, Joseph",
    editor = "Baer, H. and Belyaev, A.",
    title = "{Monopoles, duality, and string theory}",
    eprint = "hep-th/0304042",
    archivePrefix = "arXiv",
    doi = "10.1142/S0217751X0401866X",
    journal = "Int. J. Mod. Phys. A",
    volume = "19S1",
    pages = "145--156",
    year = "2004"
}

@article{Banks:2010zn,
    author = "Banks, Tom and Seiberg, Nathan",
    title = "{Symmetries and Strings in Field Theory and Gravity}",
    eprint = "1011.5120",
    archivePrefix = "arXiv",
    primaryClass = "hep-th",
    doi = "10.1103/PhysRevD.83.084019",
    journal = "Phys. Rev. D",
    volume = "83",
    pages = "084019",
    year = "2011"
}

@article{McNamara:2019rup,
    author = "McNamara, Jacob and Vafa, Cumrun",
    title = "{Cobordism Classes and the Swampland}",
    eprint = "1909.10355",
    archivePrefix = "arXiv",
    primaryClass = "hep-th",
    month = "9",
    year = "2019"
}

@article{1,
    author = "Kaidi, Justin and Tachikawa, Yuji and Yonekura, Kazuya",
    title = "{On non-supersymmetric heterotic branes}",
    eprint = "2411.04344",
    archivePrefix = "arXiv",
    primaryClass = "hep-th",
    reportNumber = "TU-1248, KYUSHU-HET-294",
    doi = "10.1007/JHEP03(2025)211",
    journal = "JHEP",
    volume = "03",
    pages = "211",
    year = "2025"
}

@article{Kaidi:2023tqo,
    author = "Kaidi, Justin and Ohmori, Kantaro and Tachikawa, Yuji and Yonekura, Kazuya",
    title = "{Nonsupersymmetric Heterotic Branes}",
    eprint = "2303.17623",
    archivePrefix = "arXiv",
    primaryClass = "hep-th",
    doi = "10.1103/PhysRevLett.131.121601",
    journal = "Phys. Rev. Lett.",
    volume = "131",
    number = "12",
    pages = "121601",
    year = "2023"
}

@article{Garfinkle:1990qj,
    author = "Garfinkle, David and Horowitz, Gary T. and Strominger, Andrew",
    title = "{Charged black holes in string theory}",
    reportNumber = "UCSB-TH-90-66",
    doi = "10.1103/PhysRevD.43.3140",
    journal = "Phys. Rev. D",
    volume = "43",
    pages = "3140",
    year = "1991",
    note = "[Erratum: Phys.Rev.D 45, 3888 (1992)]"
}

@article{3,
    author = "Fukuda, Masaki and Kobayashi, Shun K. and Watanabe, Kento and Yonekura, Kazuya",
    title = "{Black p-branes in heterotic string theory}",
    eprint = "2412.02277",
    archivePrefix = "arXiv",
    primaryClass = "hep-th",
    reportNumber = "TU-1247",
    doi = "10.1007/JHEP05(2025)043",
    journal = "JHEP",
    volume = "05",
    pages = "043",
    year = "2025"
}

@article{Hellerman:2006ff,
    author = "Hellerman, Simeon and Swanson, Ian",
    title = "{Dimension-changing exact solutions of string theory}",
    eprint = "hep-th/0612051",
    archivePrefix = "arXiv",
    doi = "10.1088/1126-6708/2007/09/096",
    journal = "JHEP",
    volume = "09",
    pages = "096",
    year = "2007"
}

@article{Hellerman:2007zz,
    author = "Hellerman, Simeon and Swanson, Ian",
    title = "{A Stable vacuum of the tachyonic E(8) string}",
    eprint = "0710.1628",
    archivePrefix = "arXiv",
    primaryClass = "hep-th",
    month = "10",
    year = "2007"
}

@article{Kaidi:2020jla,
    author = "Kaidi, Justin",
    title = "{Stable Vacua for Tachyonic Strings}",
    eprint = "2010.10521",
    archivePrefix = "arXiv",
    primaryClass = "hep-th",
    doi = "10.1103/PhysRevD.103.106026",
    journal = "Phys. Rev. D",
    volume = "103",
    number = "10",
    pages = "106026",
    year = "2021"
}

@article{BoyleSmith:2023xkd,
    author = "Boyle Smith, Philip and Lin, Ying-Hsuan and Tachikawa, Yuji and Zheng, Yunqin",
    title = "{Classification of chiral fermionic CFTs of central charge $\le$ 16}",
    eprint = "2303.16917",
    archivePrefix = "arXiv",
    primaryClass = "hep-th",
    doi = "10.21468/SciPostPhys.16.2.058",
    journal = "SciPost Phys.",
    volume = "16",
    number = "2",
    pages = "058",
    year = "2024"
}

@article{Polchinski:2005bg,
    author = "Polchinski, Joseph",
    title = "{Open heterotic strings}",
    eprint = "hep-th/0510033",
    archivePrefix = "arXiv",
    doi = "10.1088/1126-6708/2006/09/082",
    journal = "JHEP",
    volume = "09",
    pages = "082",
    year = "2006"
}

@article{Bergshoeff:2006bs,
    author = "Bergshoeff, Eric A. and Gibbons, Gary W. and Townsend, Paul K.",
    title = "{Open M5-branes}",
    eprint = "hep-th/0607193",
    archivePrefix = "arXiv",
    reportNumber = "DAMTP-2006-42, UG-06-06",
    doi = "10.1103/PhysRevLett.97.231601",
    journal = "Phys. Rev. Lett.",
    volume = "97",
    pages = "231601",
    year = "2006"
}

@article{Basile:2023knk,
    author = "Basile, Ivano and Debray, Arun and Delgado, Matilda and Montero, Miguel",
    title = "{Global anomalies {\&} bordism of non-supersymmetric strings}",
    eprint = "2310.06895",
    archivePrefix = "arXiv",
    primaryClass = "hep-th",
    reportNumber = "IFT-23-129",
    doi = "10.1007/JHEP02(2024)092",
    journal = "JHEP",
    volume = "02",
    pages = "092",
    year = "2024"
}

@article{Debray:2023rlx,
    author = "Debray, Arun",
    title = "{Bordism for the 2-group symmetries of the heterotic and CHL strings}",
    eprint = "2304.14764",
    archivePrefix = "arXiv",
    primaryClass = "math.AT",
    doi = "10.1090/conm/802/16079",
    journal = "Contemp. Math.",
    volume = "802",
    pages = "227--98",
    year = "2024"
}

@article{Kneissl:2024zox,
    author = "Kneissl, Christian",
    title = "{Spin cobordism and the gauge group of type I/heterotic string theory}",
    eprint = "2407.20333",
    archivePrefix = "arXiv",
    primaryClass = "hep-th",
    reportNumber = "MPP-2024-159",
    doi = "10.1007/JHEP01(2025)181",
    journal = "JHEP",
    volume = "01",
    pages = "181",
    year = "2025"
}

@article{Montero:2020icj,
    author = "Montero, Miguel and Vafa, Cumrun",
    title = "{Cobordism Conjecture, Anomalies, and the String Lamppost Principle}",
    eprint = "2008.11729",
    archivePrefix = "arXiv",
    primaryClass = "hep-th",
    doi = "10.1007/JHEP01(2021)063",
    journal = "JHEP",
    volume = "01",
    pages = "063",
    year = "2021"
}

@article{5,
    author = "Gubser, Steven S.",
    title = "{Colorful horizons with charge in anti-de Sitter space}",
    eprint = "0803.3483",
    archivePrefix = "arXiv",
    primaryClass = "hep-th",
    reportNumber = "PUPT-2264",
    doi = "10.1103/PhysRevLett.101.191601",
    journal = "Phys. Rev. Lett.",
    volume = "101",
    pages = "191601",
    year = "2008"
}

@article{6,
    author = "Witten, Edward",
    title = "{On string theory and black holes}",
    reportNumber = "IASSNS-HEP-91-12",
    doi = "10.1103/PhysRevD.44.314",
    journal = "Phys. Rev. D",
    volume = "44",
    pages = "314--324",
    year = "1991"
}

@article{McGuigan:1991qp,
    author = "McGuigan, Michael D. and Nappi, Chiara R. and Yost, Scott A.",
    title = "{Charged black holes in two-dimensional string theory}",
    eprint = "hep-th/9111038",
    archivePrefix = "arXiv",
    reportNumber = "IASSNS-HEP-91-57",
    doi = "10.1016/0550-3213(92)90039-E",
    journal = "Nucl. Phys. B",
    volume = "375",
    pages = "421--450",
    year = "1992"
}

@article{Gibbons:1992rh,
    author = "Gibbons, G. W. and Perry, M. J.",
    title = "{The Physics of 2-D stringy space-times}",
    eprint = "hep-th/9204090",
    archivePrefix = "arXiv",
    doi = "10.1142/S0218271892000161",
    journal = "Int. J. Mod. Phys. D",
    volume = "1",
    pages = "335--354",
    year = "1992"
}

@article{4,
    author = "Hartnoll, Sean A. and Herzog, Christopher P. and Horowitz, Gary T.",
    title = "{Holographic Superconductors}",
    eprint = "0810.1563",
    archivePrefix = "arXiv",
    primaryClass = "hep-th",
    doi = "10.1088/1126-6708/2008/12/015",
    journal = "JHEP",
    volume = "12",
    pages = "015",
    year = "2008"
}

@article{Gubser:2008px,
    author = "Gubser, Steven S.",
    title = "{Breaking an Abelian gauge symmetry near a black hole horizon}",
    eprint = "0801.2977",
    archivePrefix = "arXiv",
    primaryClass = "hep-th",
    reportNumber = "PUPT-2255",
    doi = "10.1103/PhysRevD.78.065034",
    journal = "Phys. Rev. D",
    volume = "78",
    pages = "065034",
    year = "2008"
}

@article{Hartnoll:2008vx,
    author = "Hartnoll, Sean A. and Herzog, Christopher P. and Horowitz, Gary T.",
    title = "{Building a Holographic Superconductor}",
    eprint = "0803.3295",
    archivePrefix = "arXiv",
    primaryClass = "hep-th",
    reportNumber = "NSF-KITP-08-38, PUPT-2261",
    doi = "10.1103/PhysRevLett.101.031601",
    journal = "Phys. Rev. Lett.",
    volume = "101",
    pages = "031601",
    year = "2008"
}

@article{Albash:2008eh,
    author = "Albash, Tameem and Johnson, Clifford V.",
    title = "{A Holographic Superconductor in an External Magnetic Field}",
    eprint = "0804.3466",
    archivePrefix = "arXiv",
    primaryClass = "hep-th",
    doi = "10.1088/1126-6708/2008/09/121",
    journal = "JHEP",
    volume = "09",
    pages = "121",
    year = "2008"
}

@article{Gubser:2008wv,
    author = "Gubser, Steven S. and Pufu, Silviu S.",
    title = "{The Gravity dual of a p-wave superconductor}",
    eprint = "0805.2960",
    archivePrefix = "arXiv",
    primaryClass = "hep-th",
    reportNumber = "PUPT-2270",
    doi = "10.1088/1126-6708/2008/11/033",
    journal = "JHEP",
    volume = "11",
    pages = "033",
    year = "2008"
}

@article{Breitenlohner:1982jf,
    author = "Breitenlohner, Peter and Freedman, Daniel Z.",
    title = "{Stability in Gauged Extended Supergravity}",
    reportNumber = "Print-82-0500 (MIT)",
    doi = "10.1016/0003-4916(82)90116-6",
    journal = "Annals Phys.",
    volume = "144",
    pages = "249",
    year = "1982"
}

@article{Horowitz:1993jc,
    author = "Horowitz, Gary T. and Welch, Dean L.",
    title = "{Exact three-dimensional black holes in string theory}",
    eprint = "hep-th/9302126",
    archivePrefix = "arXiv",
    reportNumber = "NSF-ITP-93-21",
    doi = "10.1103/PhysRevLett.71.328",
    journal = "Phys. Rev. Lett.",
    volume = "71",
    pages = "328--331",
    year = "1993"
}

@inproceedings{Horowitz:1992jp,
    author = "Horowitz, Gary T.",
    title = "{The dark side of string theory: Black holes and black strings.}",
    eprint = "hep-th/9210119",
    archivePrefix = "arXiv",
    reportNumber = "UCSBTH-92-32",
    month = "10",
    year = "1992"
}

@article{Abbott:2008cd,
    author = "Abbott, Michael C. and Lowe, David A.",
    title = "{Six-Dimensional Yang Black Holes in Dilaton Gravity}",
    eprint = "0802.2976",
    archivePrefix = "arXiv",
    primaryClass = "hep-th",
    reportNumber = "BROWN-HET-1493",
    doi = "10.1016/j.physletb.2008.04.064",
    journal = "Phys. Lett. B",
    volume = "664",
    pages = "214--218",
    year = "2008"
}

@article{2,
    author = "Horowitz, Gary T. and Strominger, Andrew",
    title = "{Black strings and P-branes}",
    reportNumber = "UCSBTH-91-06",
    doi = "10.1016/0550-3213(91)90440-9",
    journal = "Nucl. Phys. B",
    volume = "360",
    pages = "197--209",
    year = "1991"
}

@article{Maldacena:1997re,
    author = "Maldacena, Juan Martin",
    title = "{The Large $N$ limit of superconformal field theories and supergravity}",
    eprint = "hep-th/9711200",
    archivePrefix = "arXiv",
    reportNumber = "HUTP-97-A097, HUTP-98-A097",
    doi = "10.4310/ATMP.1998.v2.n2.a1",
    journal = "Adv. Theor. Math. Phys.",
    volume = "2",
    pages = "231--252",
    year = "1998"
}

@article{Witten:1998qj,
    author = "Witten, Edward",
    title = "{Anti de Sitter space and holography}",
    eprint = "hep-th/9802150",
    archivePrefix = "arXiv",
    reportNumber = "IASSNS-HEP-98-15",
    doi = "10.4310/ATMP.1998.v2.n2.a2",
    journal = "Adv. Theor. Math. Phys.",
    volume = "2",
    pages = "253--291",
    year = "1998"
}

@article{Giveon:2004zz,
    author = "Giveon, Amit and Sever, Amit",
    title = "{Strings in a 2-d extremal black hole}",
    eprint = "hep-th/0412294",
    archivePrefix = "arXiv",
    doi = "10.1088/1126-6708/2005/02/065",
    journal = "JHEP",
    volume = "02",
    pages = "065",
    year = "2005"
}

@article{Perry:1993ry,
    author = "Perry, Malcolm J. and Teo, Edward",
    title = "{Nonsingularity of the exact two-dimensional string black hole}",
    eprint = "hep-th/9302037",
    archivePrefix = "arXiv",
    reportNumber = "DAMTP-R-93-1",
    doi = "10.1103/PhysRevLett.70.2669",
    journal = "Phys. Rev. Lett.",
    volume = "70",
    pages = "2669--2672",
    year = "1993"
}

@article{Yi:1993gh,
    author = "Yi, Piljin",
    title = "{Nonsingular 2-D black holes and classical string backgrounds}",
    eprint = "hep-th/9302070",
    archivePrefix = "arXiv",
    reportNumber = "CALT-68-1852",
    doi = "10.1103/PhysRevD.48.2777",
    journal = "Phys. Rev. D",
    volume = "48",
    pages = "2777--2788",
    year = "1993"
}

@article{Ahn:1992df,
    author = "Ahn, Chang-Jun and Kwon, Ohjong and Park, Young-Jai and Kim, Kee-Yong and Kim, Yongduk",
    title = "{Stability of two-dimensional stringy black hole}",
    reportNumber = "SOGANG-HEP-175-92",
    doi = "10.1103/PhysRevD.47.1699",
    journal = "Phys. Rev. D",
    volume = "47",
    pages = "1699--1702",
    year = "1993"
}

@article{Li:2001ct,
    author = "Li, Xin-Zhou and Hao, Jian-Gang and Liu, Dao-Jun",
    title = "{Quasinormal modes of stringy black holes}",
    eprint = "gr-qc/0205007",
    archivePrefix = "arXiv",
    doi = "10.1016/S0370-2693(01)00437-3",
    journal = "Phys. Lett. B",
    volume = "507",
    pages = "312--316",
    year = "2001"
}

@article{Xi:2005stw,
    author = "Xi, Ping and Li, Xin-zhou",
    title = "{Object Picture of Quasinormal Modes for Stringy Black Holes}",
    eprint = "0711.4193",
    archivePrefix = "arXiv",
    primaryClass = "hep-th",
    doi = "10.1088/0256-307X/22/11/010",
    journal = "Chin. Phys. Lett.",
    volume = "22",
    pages = "2763--2765",
    year = "2005"
}

@article{Horne:1991gn,
    author = "Horne, James H. and Horowitz, Gary T.",
    title = "{Exact black string solutions in three-dimensions}",
    eprint = "hep-th/9108001",
    archivePrefix = "arXiv",
    reportNumber = "UCSBTH-91-39",
    doi = "10.1016/0550-3213(92)90536-K",
    journal = "Nucl. Phys. B",
    volume = "368",
    pages = "444--462",
    year = "1992"
}

@article{Horne:1991cn,
    author = "Horne, James H. and Horowitz, Gary T. and Steif, Alan R.",
    title = "{An Equivalence between momentum and charge in string theory}",
    eprint = "hep-th/9110065",
    archivePrefix = "arXiv",
    reportNumber = "UCSBTH-91-53",
    doi = "10.1103/PhysRevLett.68.568",
    journal = "Phys. Rev. Lett.",
    volume = "68",
    pages = "568--571",
    year = "1992"
}

@article{Banados:1992wn,
    author = "Banados, Maximo and Teitelboim, Claudio and Zanelli, Jorge",
    title = "{The Black hole in three-dimensional space-time}",
    eprint = "hep-th/9204099",
    archivePrefix = "arXiv",
    reportNumber = "PRINT-92-0151 (CHILE), IASSNS-HEP-92-29",
    doi = "10.1103/PhysRevLett.69.1849",
    journal = "Phys. Rev. Lett.",
    volume = "69",
    pages = "1849--1851",
    year = "1992"
}
\end{document}